\documentclass[11pt]{article}
\pdfoutput=1
\usepackage{jcappub}

\graphicspath{{figs/}}
\newcommand{\fig}[1]{\includegraphics[width=0.47\textwidth]{#1}}

\title{First results from the TUS orbital detector 
	in the extensive air shower mode}

\author[a]{B.A.~Khrenov,}
\author[a]{P.A.~Klimov,}
\author[a]{M.I.~Panasyuk,}
\author[a]{S.A.~Sharakin,}
\author[b,c]{L.G.~Tkachev,}
\author[a]{M.Yu.~Zotov,}
\author[b]{S.V.~Biktemerova,}
\author[d]{A.A.~Botvinko,}
\author[a]{N.P.~Chirskaya,}
\author[a]{V.E.~Eremeev,}
\author[a]{G.K.~Garipov,}
\author[b,c]{V.M.~Grebenyuk,}
\author[b]{A.A.~Grinyuk,}
\author[e]{S.~Jeong,}
\author[a]{N.N.~Kalmykov,}
\author[e]{M.~Kim,}
\author[b]{M.V.~Lavrova,}
\author[e]{J.~Lee,}
\author[f]{O.~Martinez,} 
\author[e]{I.H.~Park,} 
\author[a]{V.L.~Petrov,}
\author[f]{E.~Ponce,}
\author[d]{A.E.~Puchkov,}
\author[f]{H.~Salazar,}
\author[d]{O.A.~Saprykin,}
\author[d]{A.N.~Senkovsky,}
\author[a]{A.V.~Shirokov,}
\author[b]{A.V.~Tkachenko,}
\author[a]{I.V.~Yashin}

\affiliation[a]{Lomonosov Moscow State University, GSP-1, Leninskie
	Gory, Moscow, 119991, Russia}
\affiliation[b]{Joint Institute for Nuclear Research, Joliot-Curie, 6,
	Dubna, Moscow region, Russia, 141980}
\affiliation[c]{Dubna State University, University str., 19, Bld.1,
	Dubna, Moscow region, Russia}
\affiliation[d]{Space Regatta Consortium, ul. Lenina, 4a,
	141070 Korolev, Moscow region, Russia}
\affiliation[e]{Department of Physics and ISTS, Sungkyunkwan
	University,	Seobu-ro 2066, Suwon, \mbox{440-746} Korea}
\affiliation[f]{Benem\'{e}rita Universidad Aut\'{o}noma de Puebla,
	4 sur 104 Centro Hist\'{o}rico C.P. 72000, Puebla, Mexico}

\abstract{
	TUS (Tracking Ultraviolet Set-up), the first orbital detector of
	extreme energy cosmic rays (EECRs), those with energies above 50~EeV,
	was launched into orbit on April~28, 2016,
	as a part of the Lomonosov satellite scientific payload. 
	The main aim of the mission is to test a technique of registering
	fluorescent and Cherenkov radiation of extensive air showers
	generated by EECRs in the atmosphere with a space telescope.  
	We present preliminary results of
	its operation in a mode dedicated to registering extensive air
	showers in the period from August 16, 2016, to November 4, 2016.  
	No EECRs have been conclusively
	identified in the data yet, but the
	diversity of ultraviolet emission in the atmosphere was found to be
	unexpectedly rich.
	We discuss typical examples of data obtained with TUS and
	their possible origin.
	The data is important for obtaining more accurate
	estimates of the nocturnal ultraviolet glow of the atmosphere,
	necessary for successful development of more advanced orbital EECR
	detectors including those of the KLYPVE (\mbox{K-EUSO}) and JEM-EUSO
	missions.
}


\arxivnumber{1704.07704}

\begin{document}
\maketitle

\section{Introduction}

Extreme energy cosmic rays (EECRs), those with energies above approximately
50~EeV\footnote{$1~\mathrm{EeV}=10^{18}$~eV.}, constitute one of the
long-standing mysteries of modern astrophysics.
They were discovered more than 50 years
ago~\cite{1961PhRvL...6..485L}
but their nature and
origin still remain unclear despite the enormous efforts put into their
study, see, e.g.,~\cite{Kampert-Tinyakov-2014} for a review.
One of the main obstacles to solving the puzzle is a very low flux of
EECRs.
It suffices to say that the largest ever cosmic ray experiment, that of the Pierre
Auger Observatory, which occupies an area of more than 3000~km$^2$
in Mendoza Province, Argentina,
registered only 146 EECRs with energies above 53~EeV in nearly eight
years of operation~\cite{2015ApJ...804...15A}.  Another problem is the
incomplete coverage of the celestial sphere by existing ground
arrays: the Pierre Auger Observatory and the Telescope Array, located in
the USA, observe regions of the celestial sphere that only partially
overlap.  This complicates an analysis of arrival directions
(anisotropy) of EECRs at large scales, see, e.g.~\cite{Auger-TA-2015},
so that an experiment with a full-sky coverage becomes highly desirable.

A way to overcome both difficulties was proposed by Benson and Linsley
in early 1980's~\cite{1980BAAS...12Q.818B,Benson-Linsley-1981}.  They
suggested to use a reflector type telescope deployed on a low-orbit
satellite to register the ultraviolet (UV) photons resulting from
the fluorescence and
Cherenkov radiation emitted by ionized molecules of nitrogen excited by
charged particles of extensive air shower (EAS) cascades generated by
EECRs in the atmosphere.
In their initial proposal, the telescope was to be equipped with 
about 5000 photomultiplier tubes located at the
focal surface of a 36~m diameter mirror.
The task proved challenging in both its scientific and technical aspects
because it would require an instrument able to register, from the severe
conditions of outer space, the faint flux of fluorescence and Cherenkov
photons against the ever-changing nocturnal UV glow of the atmosphere.
As a result of these difficulties, no such projects suggested have been
implemented yet, see, e.g.,~\cite{ECRS} for a brief review.

The TUS (Tracking Ultraviolet Set-up) detector was first announced in
2001 as a pathfinder for a more advanced KLYPVE
project~\cite{2001ICRC....2..831A,2001AIPC..566...57K}.%
\footnote{%
	Remarkably, John~Linsley was one of the active members of the team.
	}
TUS inherited the optical scheme of the original design by Benson and Linsley
but with much more modest technical parameters, see
Section~\ref{sec:design}.
The D.V.~Skobeltsyn Institute of Nuclear
Physics at M.V.~Lomonosov Moscow State University led the development of
TUS in collaboration with
a number of universities and research organizations in Russia, Korea,
and Mexico~\cite{2004PAN....67.2058K,2008AdSpR..41.2079A,TUS-ecrs2012}.
TUS was launched into space on
April~28, 2016, as a part of the scientific payload of the Lomonosov
satellite~\cite{Lomonosov2013},
pioneering the exploration of EECRs from space.

The main scientific goal of TUS is to verify a technique for observing
extreme energy cosmic rays with an orbital detector. With its different
modes of operation, TUS is able to obtain information about
various kinds of transient luminous events (TLEs) in the atmosphere as
well as about the UV background.
In what follows, we briefly review the design of TUS and follow with a
presentation of the preliminary results of its operation
in its main mode, dedicated to registering EECRs.
To date, no extensive air showers have been
unambiguously identified in the data 
from observations conducted over the period
from August~16, 2016, to November~4, 2016.
Nevertheless, this is the first time the UV background of the atmosphere
has been studied in the imaging mode.
The results obtained were mostly unexpected and provide
crucial information for the development of future orbital missions such
as KLYPVE (\mbox{K-EUSO})~\cite{klypve-2015} and
JEM-EUSO~\cite{JEM-EUSO-intro,JEM-EUSO-tool}, which are likely to
constitute a
break-through in exploration of the highest energy particles from
space~\cite{Olinto-ICRC-2015,ECRS,Semikoz:2016mfg}.
A detailed analysis of the data is in progress.


\section{Design of the TUS detector}
\label{sec:design}

TUS is a component of the scientific payload of the
Lomonosov satellite (international designation MVL~300, or 2016-026A).
The satellite has a sun-synchronous orbit with an inclination
of $97^\circ\!\!.3$, a period of $\approx94$~min, and a height of about
470--500~km.
TUS operates during nocturnal segments of the orbit.

\begin{figure}[!ht]
	\centerline{\fig{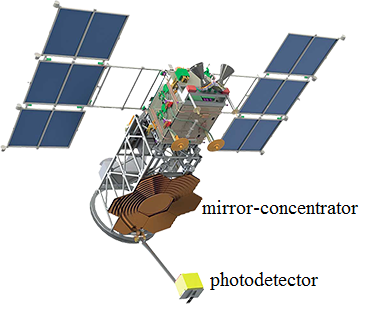}}
	\caption{Artist's depiction of the TUS detector on board the Lomonosov satellite.}
	\label{fig:MVL}
\end{figure}

The TUS detector consists of two main components: a parabolic
mirror-concentrator of the Fresnel type and a square-shaped
$16\times16$-channel
photodetector aligned to the focal plane of the mirror.  
The mirror is composed of
7 hexagonal segments of equal size, each of which is
composited from two carbon plastic layers and
strengthened by a honeycomb aluminium structure.  
The entire system has an area
of about 2~m$^2$ and a 1.5~m focal distance.
The field of view
(FOV) of the detector is $\pm4.5^\circ$, which covers an area of
approximately 80~km$\times$80~km at sea level.
The angular resolution of a single channel equals 10~mrad, which results
in a 5~km$\times$5~km area at sea level,
see~\cite{2011ICRC....3..124T,2014NIMPA.763..604G} for more details.
An artist's view of
the TUS detector on board the Lomonosov satellite is shown in
figure~\ref{fig:MVL}. 

{\sloppy
Pixels of the photodetector are Hamamatsu R1463 photomultiplier
tubes (PMTs) with a 13~mm diameter multialkali cathode.  Their quantum
efficiency is about~20\% for the 350~nm wavelength. Light guides with
square entrance apertures (15~mm$\times$15~mm) and circular outputs were
employed to uniformly fill the detector's field of view with 256 pixels.
All PMTs have black blends extending 1~cm above their light
guides as protection against side illumination. 
A 13~mm diameter, 2.5~mm thick 
UV filter is placed in front of each PMT cathode to limit the measured
wavelength to the 300--400~nm range.

}

The pixels are grouped in 16 identical photodetector modules,%
\footnote{%
	In what follows, the 16 modules form horizontal rows in figures
	presenting snapshots of the focal plane.}
each of which has its own digital data processing system for the
first-level trigger, based on a Xilinx field-programmable gate array
(FPGA), and a high voltage power supply, controlled by the FPGA. The
central processor board gathers information from all modules, controls
their operation, and implements the second-level trigger algorithm.
More details on the design of TUS and its photodetector operation can
be found in~\cite{SSR,2011ICRC...11..272T}.

The TUS electronics can operate in four modes intended for detecting
various fast optical phenomena in the atmosphere on different time
scales.
The main mode is aimed at registering EASs generated by extreme energy
cosmic rays and has a time sampling window of 0.8~$\mu$s.  Our present
study focuses on results obtained in this mode of operation. 
Three other modes are also available.
Time sampling windows of 25.6~$\mu$s and 0.4~ms are utilized for
studying transient luminous events of different kinds, and a window of
6.6~ms is available for detecting micro-meteors and possibly space
debris.
Data records for all TUS events include 256 discrete waveforms, one for
each channel, and every waveform contains analog-to-digital converter
(ADC) counts for 256 time frames.

The TUS trigger system employs a simple algorithm that
consists of two levels. The first level trigger
decision is based on a comparison of a sum of ADC counts calculated for
each pixel during 16 time frames with a threshold level.
The total duration of 16 frames (12.8~$\mu$s) corresponds to the time
necessary for the development of a nearly horizontal EAS within a pixel
FOV (approximately 5~km$\times$5~km).
In its turn, the threshold level depends on the mean value of the
background radiation within the FOV of a pixel, calculated during
100~ms.

A contiguity trigger is implemented at the second level.  The algorithm
selects pixels that satisfy the first level trigger requirements and
that are adjacent both temporally and in the focal plane.  The trigger
employs an additional parameter, a so-called adjacency length~$L$, i.e.,
the number of neighboring pixels sequentially triggered at the first
level.  A number of values in the range from~3 to~6 were tested during
the first weeks of TUS operation, and $L=6$ was chosen for the period of
data acquisition discussed below as a compromise between the aim to
register weak signals of EASs and the need to suppress too high trigger
rate due to random fluctuations of the background radiation that takes
place for low values of~$L$.

Two processes in the photodetector electronics affect event selection by
TUS.  These are the trigger algorithm itself and the PMT gain control.
During normal operation, the detector measures the UV background level
and adjusts the sensitivity of the PMTs to avoid their saturation under
conditions of increased UV radiation intensity, for example, during
moonlit nights and above areas of auroral and thunderstorm activity,
large cities, etc.  This leads to a diminished sensitivity above these
regions and a higher trigger threshold.  The high voltage correction
occurs once every 100~ms to ensure a constant anode current during EAS
waveform measurements.  Information about high voltage for all PMT
modules is included in each data record.  Due to certain limitations of
the Lomonosov hardware, the rate of TUS events recorded by the so-called
Information Unit of the satellite does not exceed one event per
approximately a minute.

\begin{figure}[!ht]
	\centerline{\fig{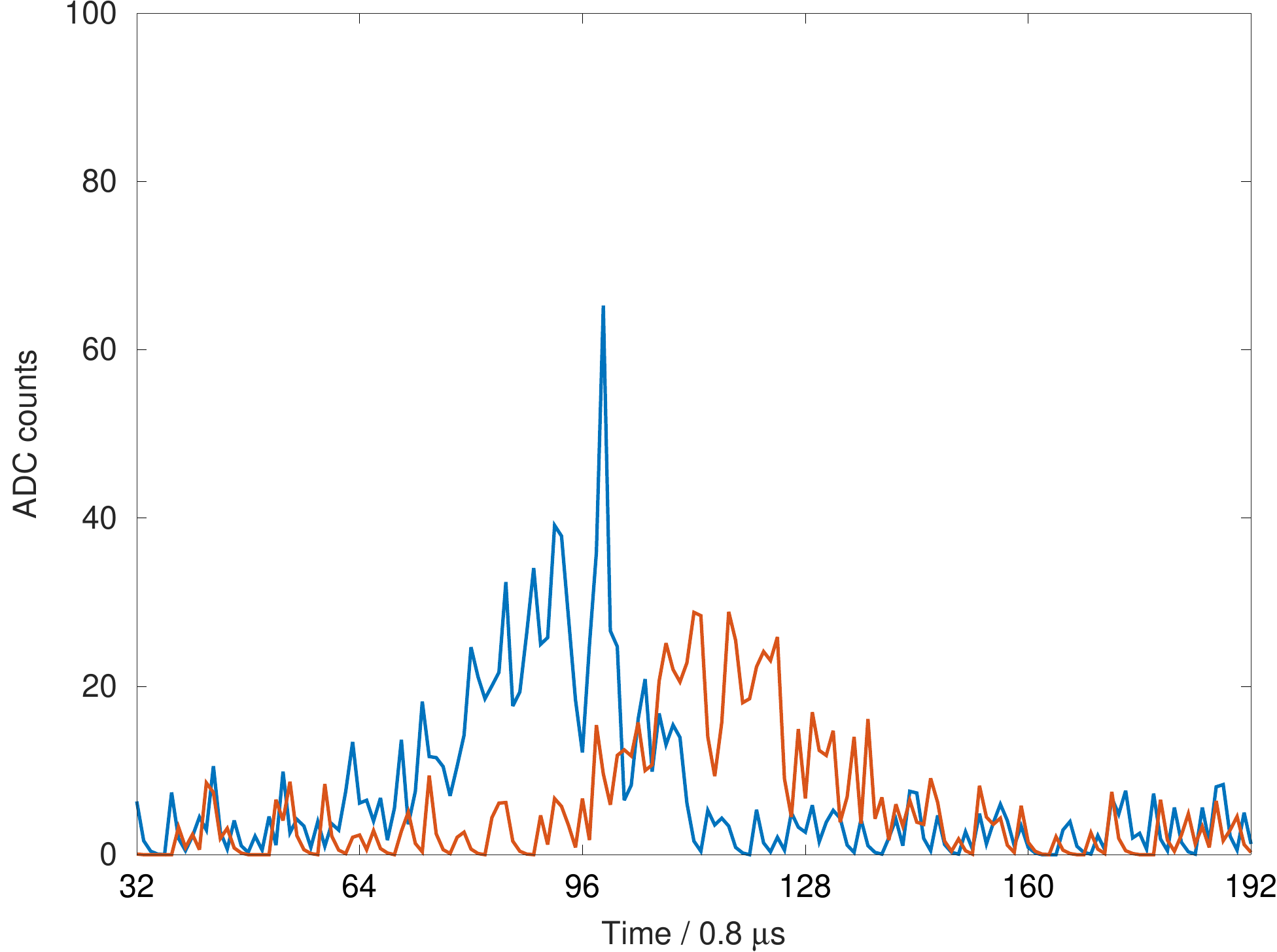}}
	\caption{Example of simulated waveforms for a nearly horizontal 100~EeV proton.}
	\label{fig:sim_waveforms}
\end{figure}

\begin{figure}[!ht]
	\centerline{\fig{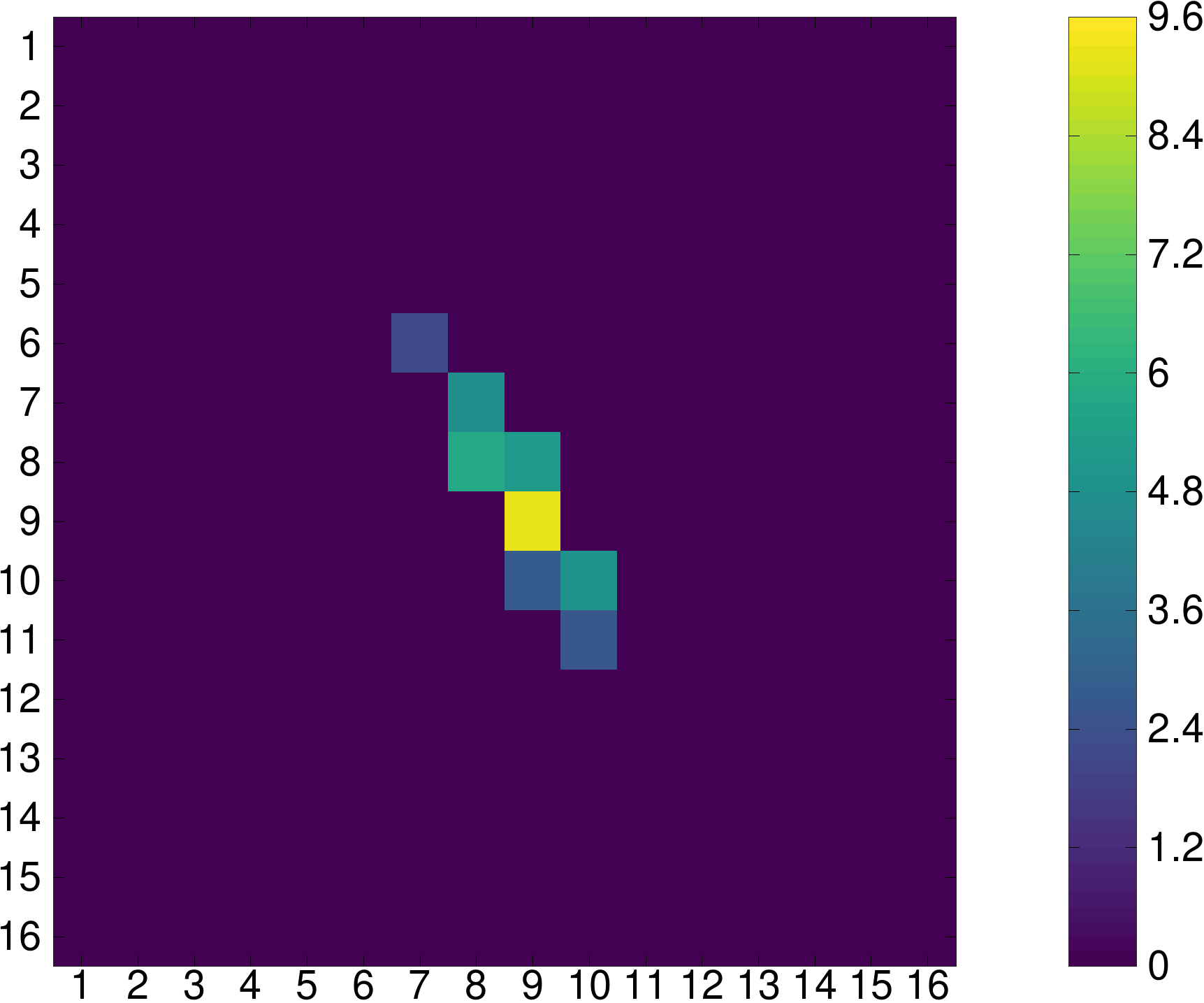}}
	\caption{Distribution of the signal from a simulated 100~EeV proton.
		Colours denote the ratio~$R$, defined in Eq.~(\ref{eq:R}).
		Only pixels with $R>2$ are selected.}
	\label{fig:sim_track}
\end{figure}

Intensive simulations of the EAS development and response of TUS have
been performed before the launch.  They employed the ESAF
framework~\cite{ESAF2010} and a dedicated TUSSIM program developed at
Joint Institute for Nuclear Research,
Dubna~\cite{2013JPhCS.409a2105G,sim-icrc2015,TUS-sim-ApP}.
Figures~\ref{fig:sim_waveforms} and~\ref{fig:sim_track} illustrate a
few important results of the simulations.
Specifically, figure~\ref{fig:sim_waveforms} demonstrates how the waveforms of
two adjacent pixels might look in the case of a nearly horizontal EAS
generated
by a proton with an energy of $\sim100$~EeV. Here, a minimal UV background
of the order of $3\cdot10^7$~ph~sr$^{-1}$~s$^{-1}$~cm$^{-2}$ is assumed.
Figure~\ref{fig:sim_track}, adopted from~\cite{TUS-expastron},
shows how the signal from an EAS can be distributed in the
photodetector. The result was obtained for a 100~EeV primary proton
arriving in the atmosphere at a zenith angle of $75^\circ$, and
a UV background glow of the order of
$5\cdot10^7$~ph~sr$^{-1}$~s$^{-1}$~cm$^{-2}$.
The signal is shown as a ratio 
\begin{equation}
	R=(A-\langle A\rangle)/\text{RMS}(A),
	\label{eq:R}
\end{equation}
where $\langle A\rangle$ is the mean value of ADC counts~$A$, and
$\text{RMS}(A)$ is the root mean square of~$A$.

The energy threshold for TUS was determined to be approximately 70~EeV
during moonless nights, and the instrument should be able to register
several EECRs above the threshold in 5 years of continuous
operation~\cite{TUS-sim-ApP} assuming the energy spectrum obtained with
the Pierre Auger Observatory for highly inclined
EAS~\cite{Auger-spectrum-2015}.  The energy threshold only weakly
depends on the value of adjacency length employed in the second level
trigger~\cite{TUS-sim-ApP}.

\section{Results}

In what follows, we present preliminary results of an analysis of
data obtained with TUS in the EAS mode of operation from August~16,
2016, to
November~4, 2016, during the interval between two software updates.
In total, 17345 events registered during the night time segments of orbits were
considered.
They were tentatively divided into four main groups
on the basis of the temporal structure of waveforms.
These groups consist of:
\begin{itemize}
	\item events with noise-like waveforms and
		approximately constant mean values of ADC counts,
	\item instant flashes with linear tracks across the focal
		plane (rise time $\sim1~\mu$s),
	\item slow spatially extended flashes
		(typical rise time $\gtrsim100~\mu$s),
	\item events with complex spatio-temporal dynamics of waveforms
		and images in the focal plane.
\end{itemize}
Each of the groups will be discussed below.

\subsection{Events with noise-like waveforms}

More than 80\% of events registered thus far by TUS in its main mode of
operation have noise-like waveforms with ADC counts
of all PMTs fluctuating around average values that are close to
each other when rescaled according to their individual PMT gains.
The average value of ADC counts may be higher or lower depending on
the current level of background radiation but the shape of
waveforms remains qualitatively similar.

As a rule, the illumination of the focal plane is approximately uniform for
noise-like events, and triggering seems to result from certain 
random fluctuations of the UV background.  Currently, such events
are utilized for estimating the UV background, which varies according to
the phase and
relative position of the Moon, the cloud coverage and other conditions.
The measurements from these events are also used for preparing
time-dependent maps of the UV background.  

\begin{figure}[!ht]
	\centerline{\fig{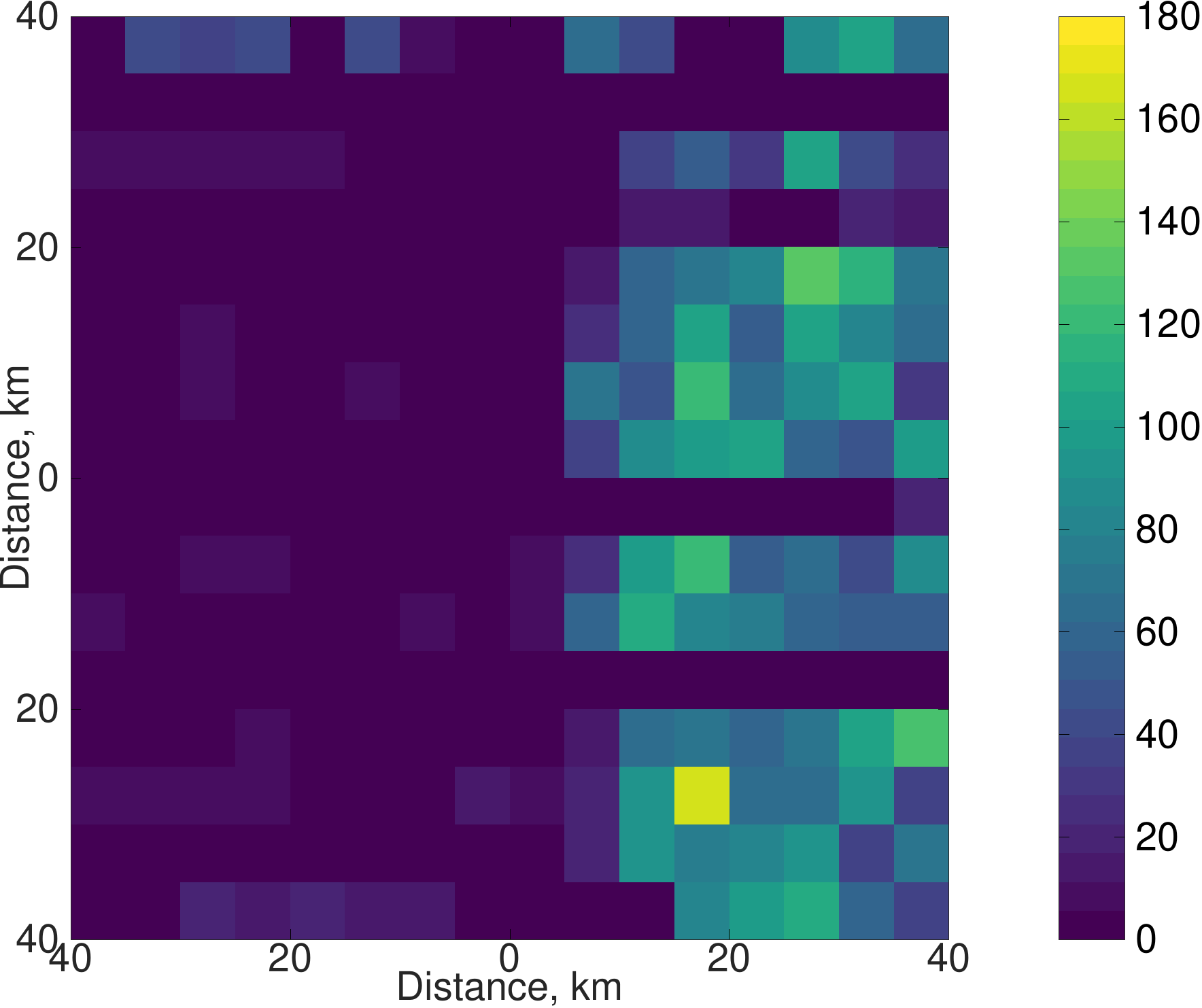}}
	\caption{Snapshot of the focal plane for
		an event registered on September~18, 2016, above the Caribbean.
		Colors denote the ADC counts.
		Here and in all subsequent figures, other than figure~\ref{fig:track},
		the numbers to the left and beneath the snapshots indicate
		approximate distances from the center of the field of view at
		ground level.
		Geographic North is approximately $11.2^\circ$ counterclockwise
		from the top of the focal plane, East is respectively to the right.
		}
	\label{fig:moon}
\end{figure}

\begin{figure}[!ht]
	\centering
	\includegraphics[width=\textwidth]{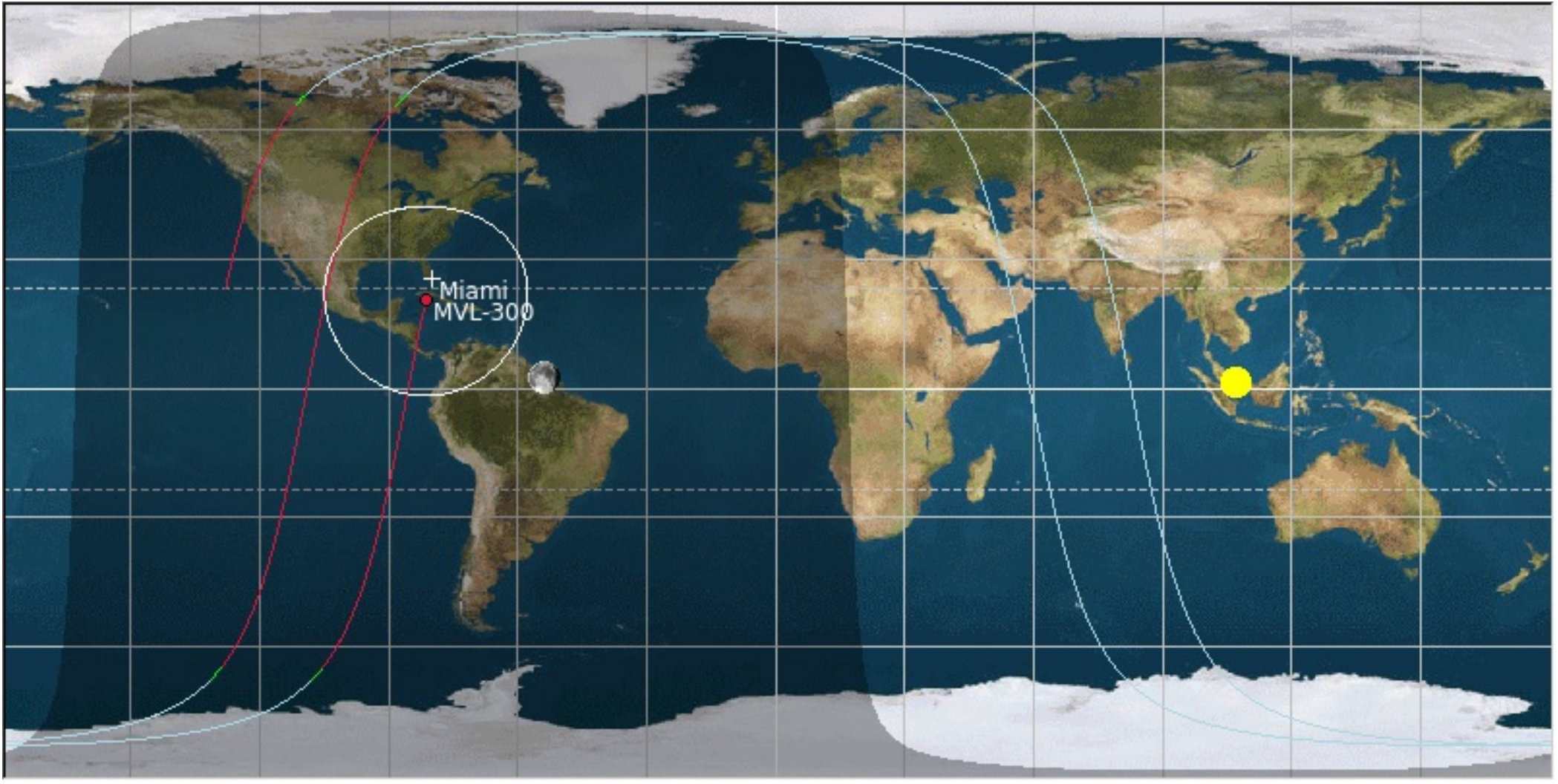}
	\caption{Relative positions of the Lomonosov satellite (MVL-300) and
		the Moon at the moment of the event shown in figure~\ref{fig:moon}.
		The figure was made with the PreviSat program
		(\texttt{http://previsat.sourceforge.net}).
		}
	\label{fig:tus-moon}
\end{figure}

Sometimes, the illumination of the focal plane is non-uniform,
displaying a
large brighter region while the waveforms
remain noise-like.  
Such conditions have been observed for numerous events registered
during nights around full moons.  An example is shown in
figure~\ref{fig:moon}.  The event was one of a series recorded
on September~18, 2016, two days after the full moon.%
\footnote{%
	Four modules with (nearly) zero ADC counts were not working properly at
	the time.}
An explanation for this kind of illumination is that, for certain
relative positions of the satellite and the Moon, the moonlight 
arrives at the focal surface directly, without previous reflection by the
atmosphere, see figure~\ref{fig:tus-moon}.
Additional light above auroral ovals or thunderstorm regions can also
contribute to non-uniform illumination of the detector's focal
plane.

\begin{figure}[!ht]
	\centerline{\fig{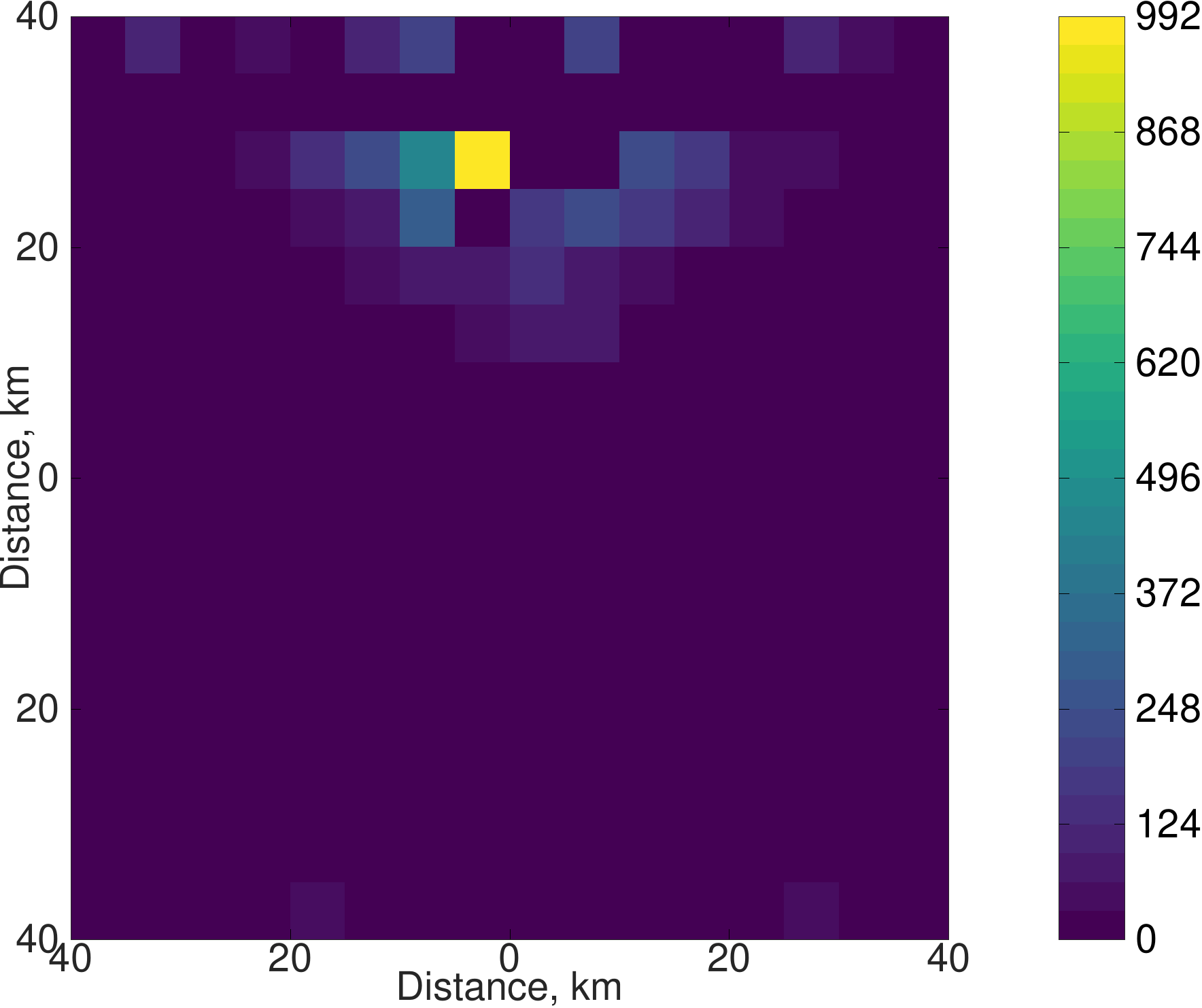}}
	\caption{Typical snapshot of the focal plane of
		an event registered on September~30, 2016, near Las Vegas, USA.
		Geographic North is approximately $12^\circ$ counterclockwise
		from the top of the focal plane, East is respectively to the right.
		}
	\label{fig:las_vegas}
\end{figure}

Within the group, there is a subset of events with noise-like waveforms,
but strongly non-uniform illumination of the focal plane.  An event of
this kind, shown in figure~\ref{fig:las_vegas}, was recorded on
September~30, 2016, at 07:11:56 UTC, when the center of the FOV of TUS
was about 25~km south from Las Vegas, Nevada (USA).  One can see a strong
localized signal, in this particular case covering a part of the city.
Two other examples of events recorded above cities can be found
in~\cite{uhecr2016}.
TUS is routinely registering events of this kind but we cannot 
claim that all of them relate to city lights or other anthropogenic
sources.

\begin{figure}[!ht]
	\centerline{\fig{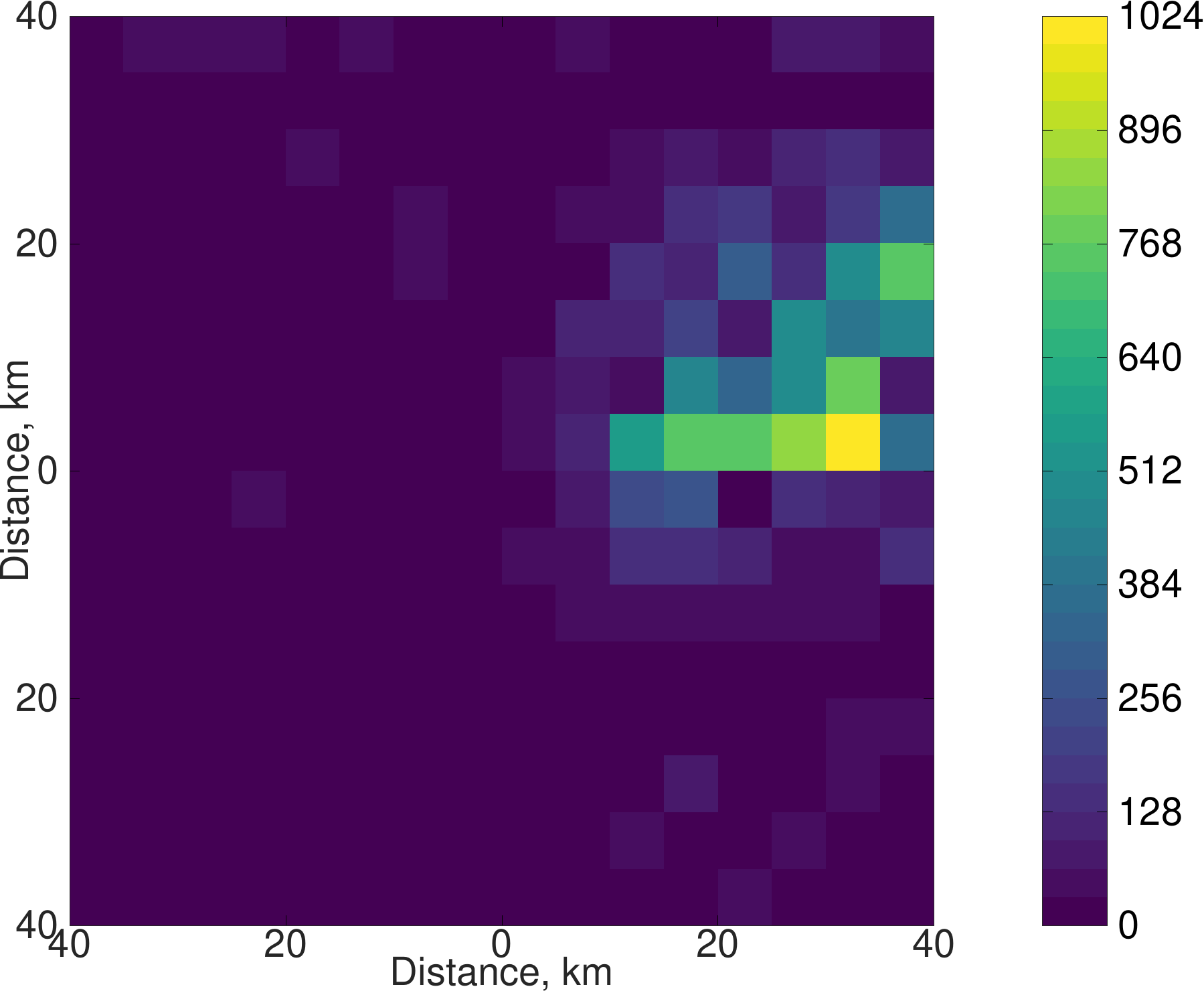}}
	\caption{Snapshot of the focal plane of
		an event registered on October~12, 2016, above Venezuela.
		}
	\label{fig:ven}
\end{figure}

In certain events, a number of pixels are saturated and sometimes,
in extreme cases of non-uniform illumination of the focal plane,
maximum ADC counts remain constant throughout the entire recording
interval.%
\footnote{%
	The maximum ADC count equals 1023, which 
	corresponds to approximately 2000 photons at a PMT pupil if the high
	voltage of the
	respective module has the maximum possible value, i.e., the module
	operates at its highest sensitivity.}
Saturated pixels together with adjacent bright ones form compact
spatially localized ``hot spots.''
Twenty events of this kind were registered within the period analyzed.
One of the events, shown in figure~\ref{fig:ven},
took place on October~12, 2016, at 03:29:49 UTC, when TUS was above
Venezuela at the $9^\circ\!\!.7$N, $63^\circ\!\!.8$W coordinates.
Remarkably, another event of this type was registered nearby,
as well as an additional one with lower ADC counts.
The region is known to have one of South America's highest levels
of anthropogenic brightness at visible
wavelengths~\cite{2016SciA....2E0377F}.

Using publicly available resources in the Internet, we found objects
like airports, power plants, and offshore platforms within the FOV of TUS
for many events of this kind.
On the other hand, we have failed to attribute these events to
atmospheric phenomena such as lightning strikes or TLEs, which
are likely to give rise to other events registered by TUS, see
Section~\ref{sec:slow_flashes} below.
Thus it is likely they are due to anthropogenic factors (for example,
LED lights or Xenon lamps) although their actual origin remains uncertain.
A detailed analysis of all events with noise-like waveforms is in
progress.

\subsection{Instant track-like flashes}

\begin{figure}[!ht]
	\includegraphics[width=.51\textwidth]{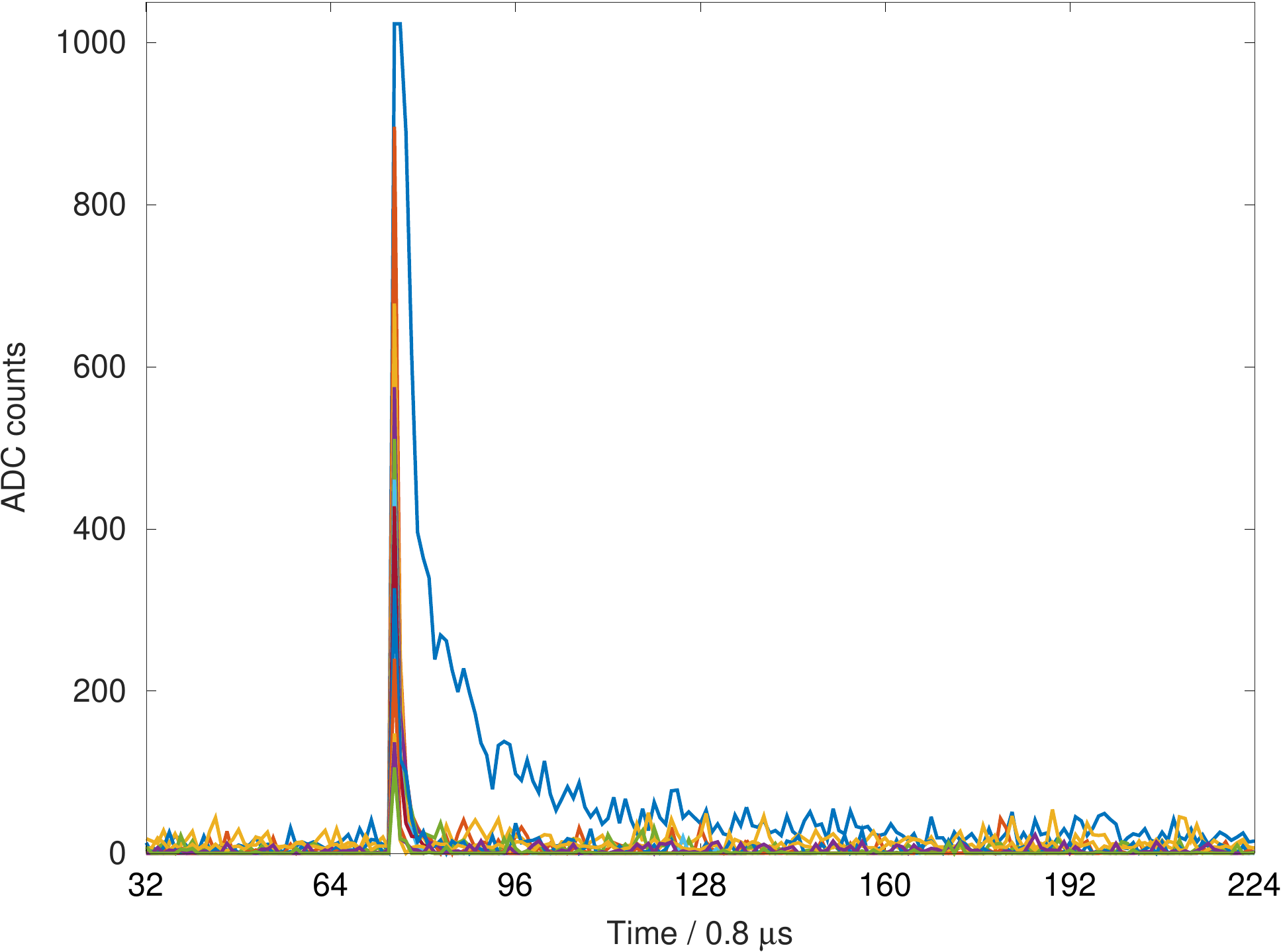}\quad
	\includegraphics[width=.47\textwidth]{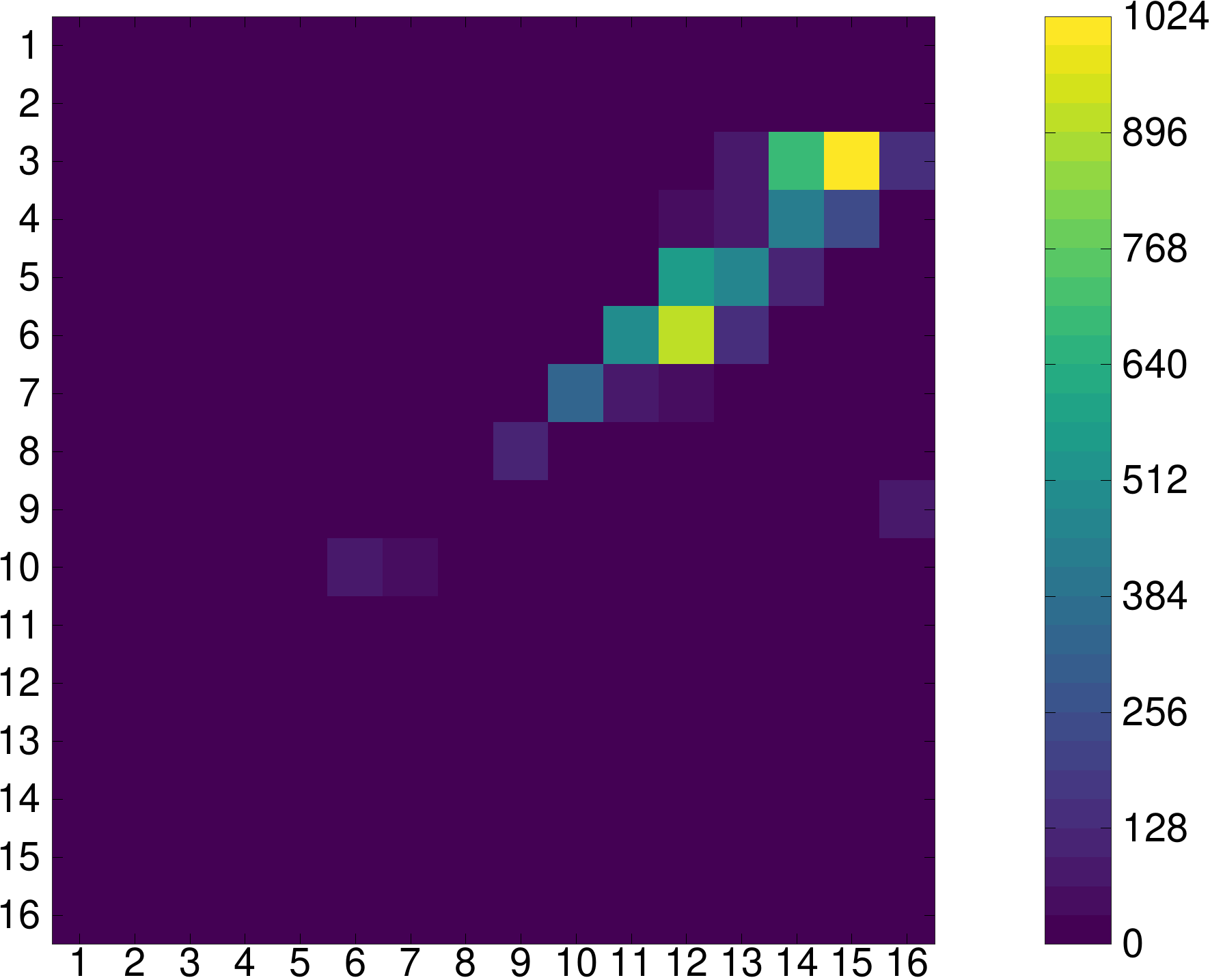}
	\caption{%
		Track-like event registered on October~25, 2016. 
		Left: waveforms of ten PMTs that demonstrated the highest ADC
		counts. Colors denote different pixels.
		Right: snapshot of the focal plane at the moment of maximum ADC counts.
		}
	\label{fig:track}
\end{figure}

One of the first phenomena that attracted our attention from the very
beginning of the TUS operation in space were exceedingly brief (i.e.,
occurring within a single or, rarely, two time frames) and, as a rule,
intense flashes that produced linear tracks across the focal surface.
In what follows, we shall call them ``track-like events'' for the sake
of brevity.
Such events comprise approximately~14\% of the total, and
up to $\sim25$\% of events registered during moonless nights.
An example is shown in figure~\ref{fig:track},
where a flash occurring during one exposure frame simultaneously%
	\footnote{In this case, ``simultaneous'' indicates an exposure
	with a duration of up to one frame, which relates to the way 
	information from PMT modules is collected.}
registers across a group of PMTs in linear alignment to form a track.  
Typically, at least one of PMTs involved in such
an event is saturated.
In a few cases, the number of saturated PMTs exceeds~8.
Another feature observed in the majority of track-like events is a
signal ``tail'' that typically lasts approximately 60--70~$\mu$s in at
least one of the saturated pixels, see figure~\ref{fig:track}.
Other examples of such flashes can be found
in~\cite{tracks-izvran,ecrs2016,uhecr2016}. 

\begin{figure*}[!ht]
	\centerline{%
		\includegraphics[height=50mm]{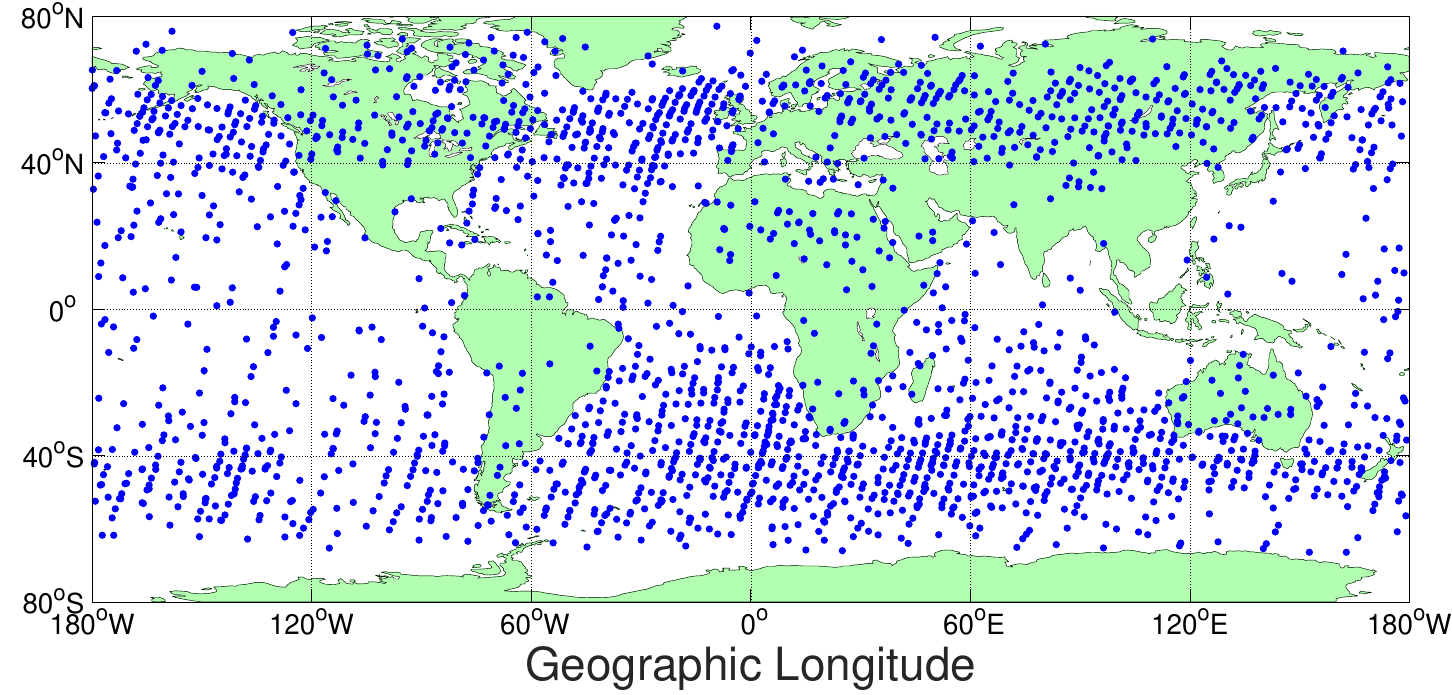}\quad
		\includegraphics[height=50mm]{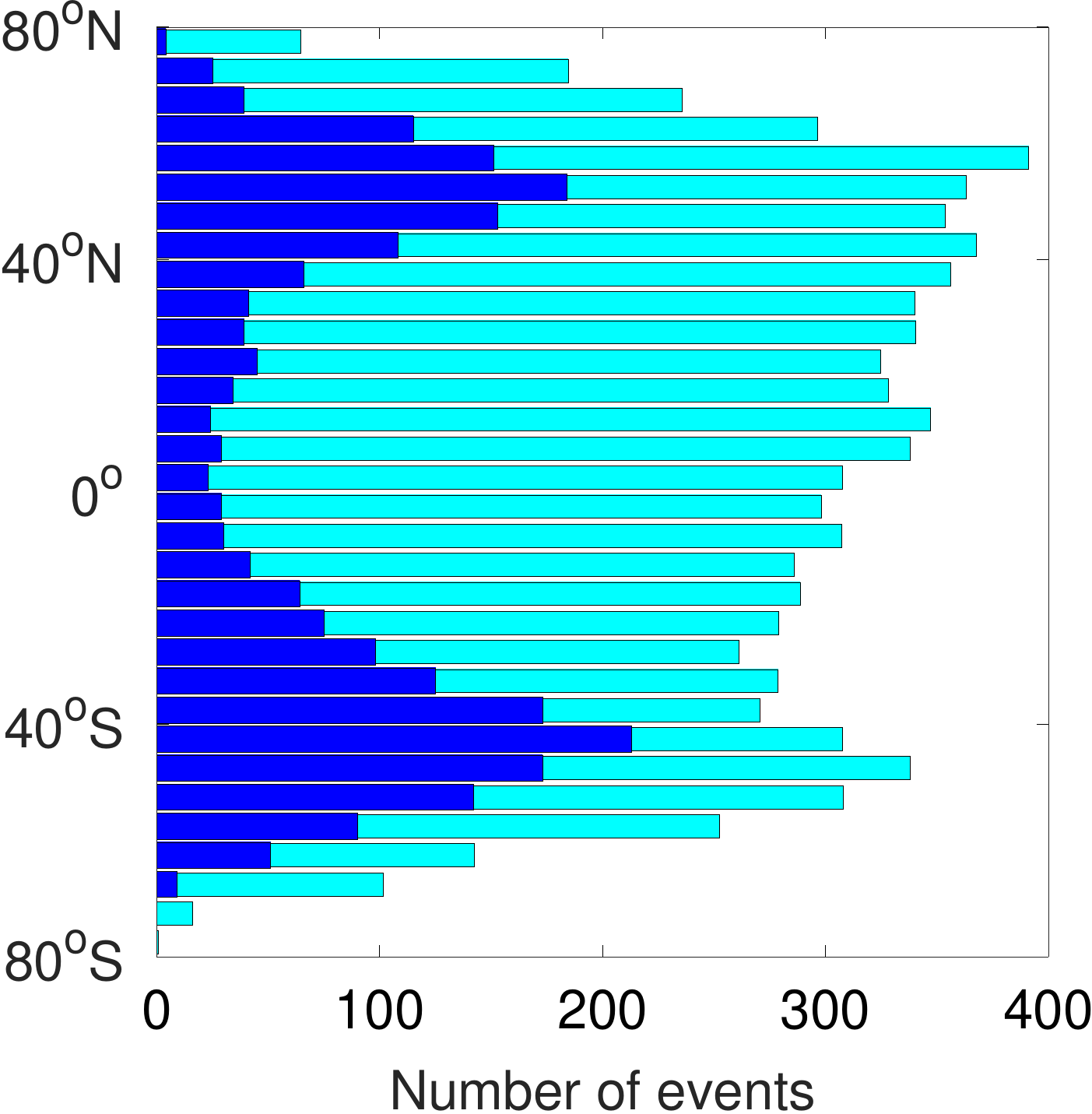}
		}
	\caption{Left: geographic distribution of 2394 track-like events over
		the Earth.
		Right: distribution of total event count (light) 
		relative to track-like events (dark) with respect to
		geographic latitude. 
		The values for the total count have been divided by~2
		for a more compact graphic representation.
	}
	\label{fig:tracks_on_earth}
\end{figure*}

Figure~\ref{fig:tracks_on_earth} shows a geographic distribution of
2394 track-like events over the Earth.%
\footnote{In 2087 cases, at least one of the PMTs was saturated.}
It is readily apparent (and confirmed by the Kolmogorov--Smirnov test) that
this mapping does not follow the distribution of all events with respect to
latitude, but has clear peaks at $50^\circ$N--$55^\circ$N and
$40^\circ$S--$45^\circ$S.
The median value of the latitude for the whole data set is
$9^\circ\!\!.4$N, and $16^\circ\!\!.5$S for the track-like events.
The distribution of all events is not symmetric with respect to the
equator due to the seasonal movement of the terminator.

A considerable population of flashes is likewise apparent within the
boundaries of the South Atlantic Anomaly, apart from a region of South
America, mostly in Brazil.  Many noise-like events characterized by a
markedly non-uniform illumination of the focal plane have been
registered in the region, likely causing a reduction in the trigger rate
of track-like events.

Distribution of track-like events versus geographic longitude is 
similar to that of the entire data set.

While interesting, these events cannot result from extensive air
showers generated by EECRs in the atmosphere because nearly horizontal
EASs should produce tracks composed of adjacent pixels that flash
throughout approximately consecutive 16 exposure frames
but not all at once.
Preliminary simulations performed using the Geant4 software
toolkit~\cite{Geant4} have revealed that protons with energies from
100--200~MeV up to a few GeV that hit the UV filters approximately
parallel to their plane can produce fluorescence and Cherenkov radiation
and result in tracks similar to those observed by TUS.

A model of the detector employed in the simulations was a matrix
of $16\times16$ cells, which consisted of circular glass filters
placed in aluminium plates with the sizes of the real detector.
Optical photons generated during interactions of a primary particle
and the material of the instrument were recorded at the bottom of each
cell, where PMTs are located.
The standard FTFP\_BERT physics list, which implements
electromagnetic and hadronic processes, was used for the simulations
that included the birth of Cherenkov photons, glass fluorescence,
absorption of photons in the glass filters and their reflection
on the boundaries of different materials.
The specific light yield of fluorescence in glass was assumed to be
300 photons/MeV at wavelengths 300--400~nm.
The refractive index of the UFS-1 glass used in filters equals~1.54.

The simulations demonstrated that a proton that hits the glass filters
approximately parallel to their plane can cause a trigger in a PMT
located near the boundary of the photodetector beginning from energies 
$\gtrsim100$~MeV. Up to six lined-up PMTs can be triggered by a 200~MeV
proton. At energies $\gtrsim400$~MeV, a proton can cause a
strong signal in sixteen and more PMTs located near a line,
with the number of photoelectrons produced that is in agreement
with ADC counts registered in the experiment.
More details on the simulations can be found in~\cite{tracks-izvran}.

The same model of the detector was used to check if electrons of the
inner Van~Allen belt can be a source of track-like events. Electrons
with energies up to 10~MeV
were considered.
It was found that an electron that hits a UV filter normally to
its plane produces a signal not exceeding the UV background level of the
nocturnal atmosphere.
On the other hand, electrons with energies less than 2--10~MeV
approaching the focal surface approximately parallel to its plane, cannot
penetrate through the aluminium box of the photodetector.
Thus electrons of this origin are unlikely to be
a source of track-like events.



It is important to note that Cherenkov light, either reflected from
Earth or generated within the mirror medium by an upward-going particle
can generate events with a strong signal and a compact spot on the
focal plane~\cite{2011BRASP..75..381S,2014AdSpR..53.1515B}. It would be
important to register an event of this kind but their duration is too
short to satisfy the second level trigger conditions with its current
settings.

\subsection{Slow spatially extended flashes}
\label{sec:slow_flashes}

\begin{figure}[!ht]
	\centerline{\fig{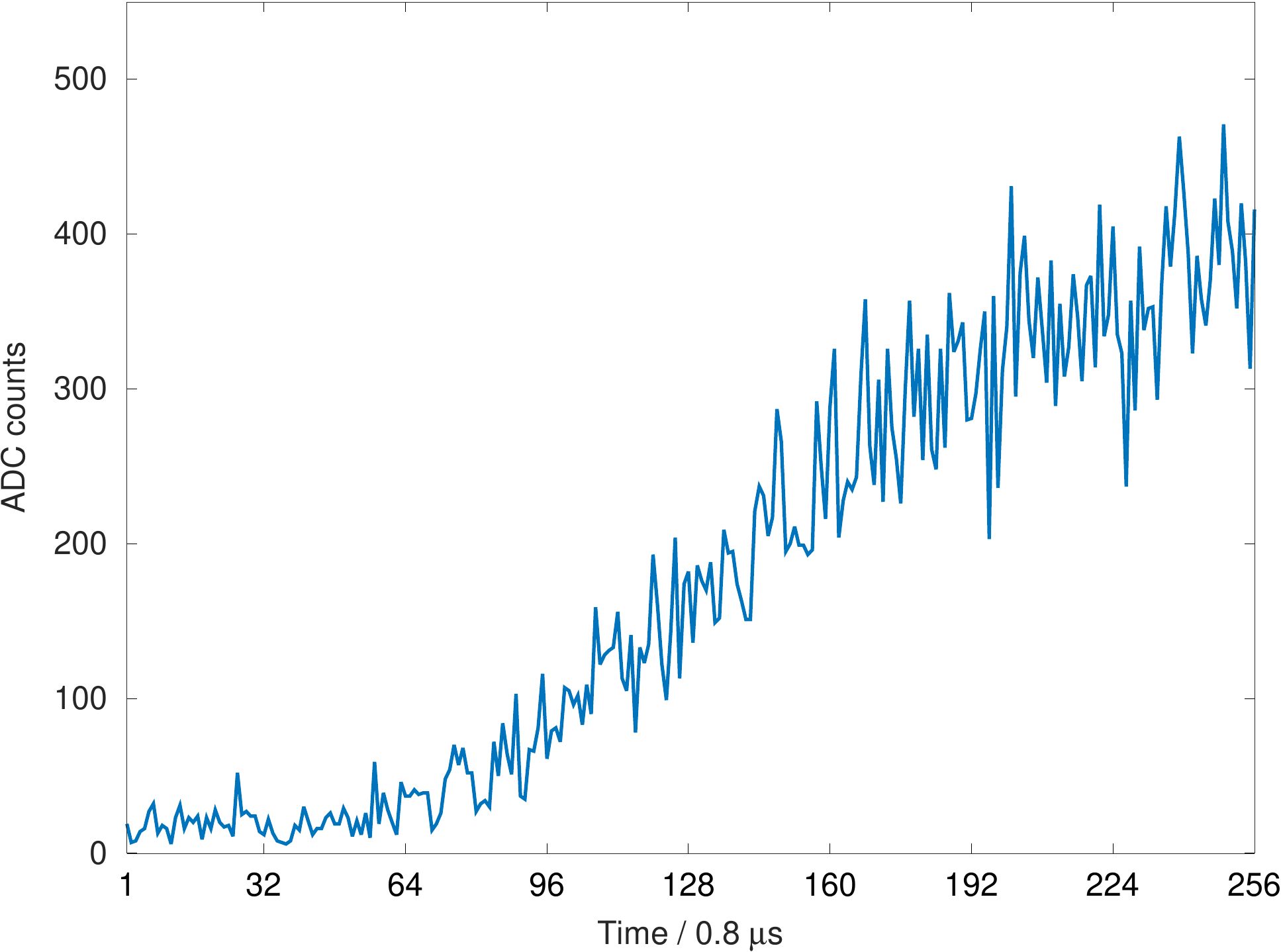}}
	\caption{Example of a waveform of a slow flash for one channel.}
	\label{fig:slow_flash_adc}
\end{figure}

Another distinct group of events consists of flashes with ADC counts
increasing steadily for $\gtrsim100~\mu$s, modulated by random
fluctuations of the signal, as shown in figure~\ref{fig:slow_flash_adc}
for a single channel. 
We shall call them ``slow flashes.''
Such a flash  typically evolves simultaneously across the majority of
pixels, producing a nearly uniform illumination of the focal plane.
In most cases, ADC counts continue to grow until the completion of a
record.  

\begin{figure*}[!ht]
	\centerline{\includegraphics[height=59mm]{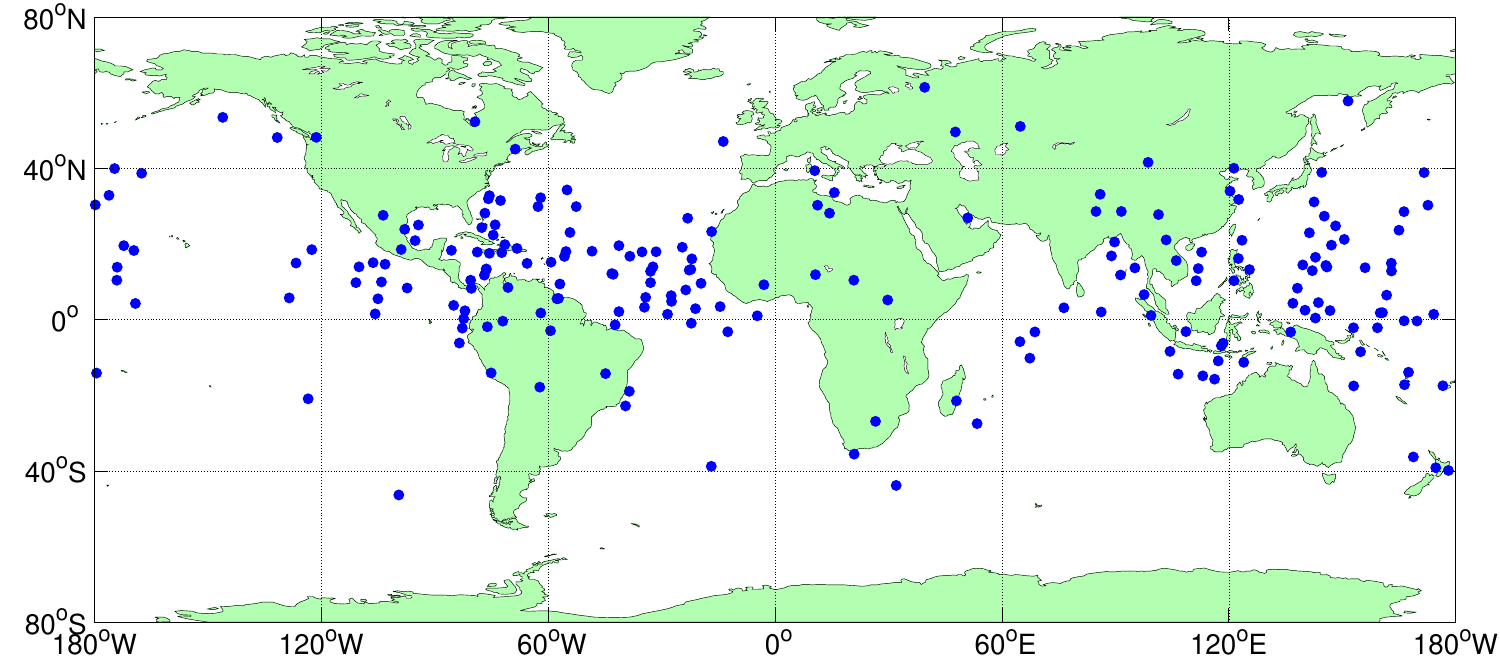}}
	\caption{Geographic distribution of 207 slow flashes.}
	\label{fig:flashes_on_earth}
\end{figure*}

The geographic distribution of slow flashes
is presented in figure~\ref{fig:flashes_on_earth}.
The selection of events shown here is conservative in the sense that
we only took flashes in which at least 25\% of all functioning PMTs
demonstrated a more than $10\sigma$ excess over the baseline signal
level near the maximum.
It is clearly seen that the majority of these TUS events are located in regions
with known high lightning flash rates.\footnote{%
	See, for example,	\texttt{http://www.sos.noaa.gov/Datasets/dataset.php?id=6}
	}
We compared a sample of 71 slow flashes registered from August~16, 2016,
to September~19, 2016, with data from the World-Wide Lightning Location Network
(WWLLN)\footnote{\texttt{http://www.wwlln.net/}} looking for
``companion'' lightnings within a wide range of time windows and at
angular distances up to~$24^\circ$ from the respective TUS event, which
roughly corresponds to the maximum distance from which a ray of light
tangent to the Earth is visible from the Lomonosov orbit.

What came as a surprise was the fact that for the time window of~$\pm1$~s,
which matches the accuracy of the trigger time stamps,
the overwhelming majority of companion lightnings were registered
at distances $>400$~km from the position of a slow flash.
In the particular case shown in figure~\ref{fig:slow_flash_adc}, two
strong lightning strikes were registered by the WWLLN at distances of
$\sim600$~km from the flash and seven weaker strikes at $\sim550$~km.
For 29 events, lightnings were found at distances $\lesssim800$~km
from the center of the TUS field of view and at distances of up to
$\sim1600$~km for 50 events.
We also checked much longer time intervals in order to figure out if
there were thunderstorms at the respective regions.
It was found that thunderstorms were registered within 30~min
at distances $\lesssim400$~km from 24 slow flashes.
The lack of lightnings strikes registered shortly before the slow flashes
might relate to the incomplete effectiveness of the WWLLN in certain
regions of the world.

The simplest way to explain this kind of flashes is that the detector
registers UV~light uniformly distributed in all directions from the location
of a lightning strike (or flash) and then partially scattered by the
surface of the mirror.

\begin{figure}[!ht]
	\centerline{\fig{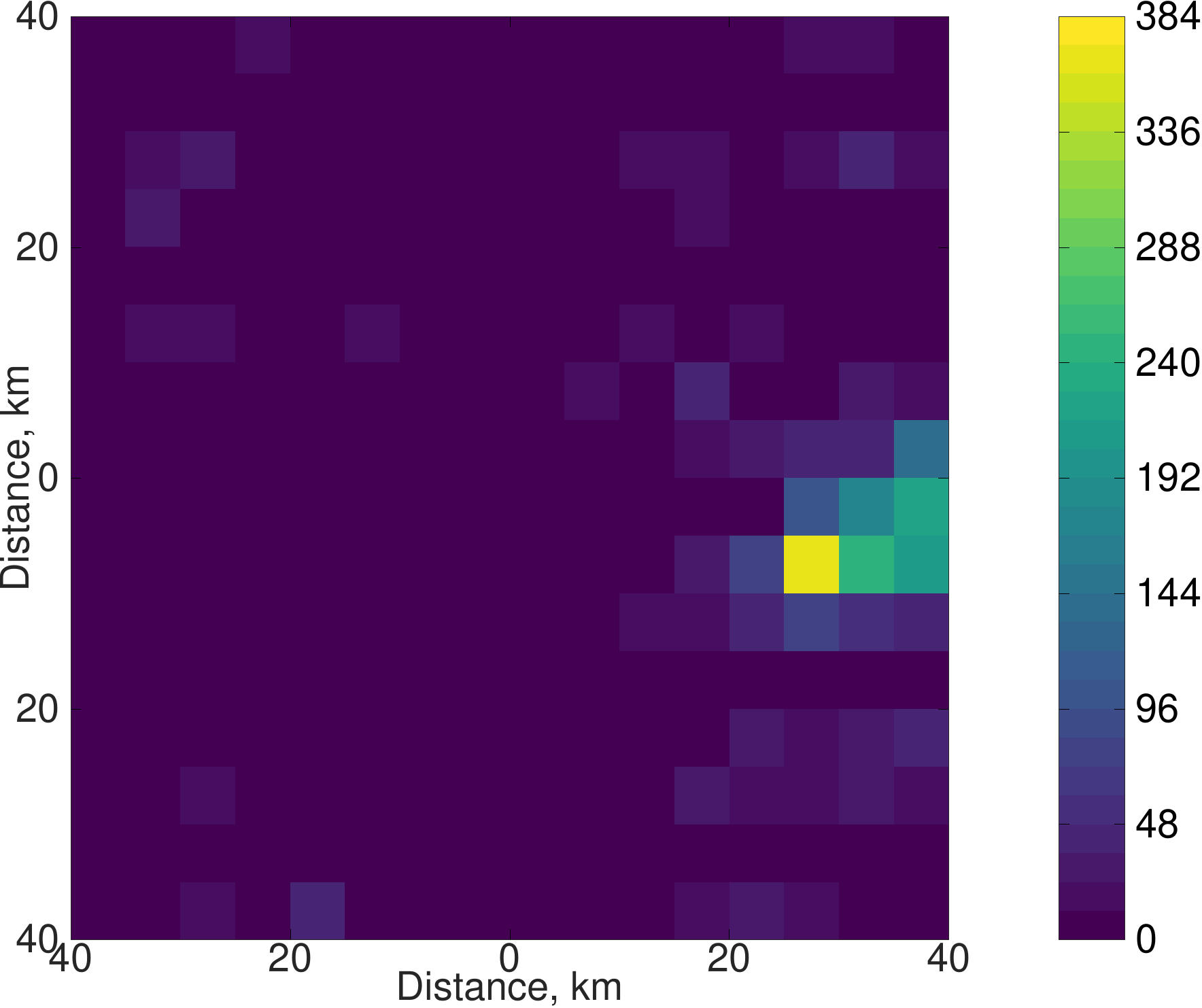}}
	\caption{Snapshot of the focal plane for
		a non-uniform slow flash registered on September~24, 2016,
		over Uganda.
		The snapshot corresponds to the frame with the maximum
		ADC count.
		}
	\label{fig:lightning}
\end{figure}

In a few cases, the illumination of the focal plane by a slow flash
is strongly non-uniform. For example, an event shown in figure~\ref{fig:lightning}
was recorded on September~24, 2016, at 21:01:46 UTC above Uganda
($1^\circ\!\!.7$N, $32^\circ\!\!.3$W).
Twenty-two PMTs demonstrated a growth of ADC counts exceeding
the baseline signal level by more than~$10\sigma$.
Two lightning strikes within a time interval of $\pm1$~s were registered by
the Vaisala Global Lightning Dataset GLD360~\cite{vaisala1,vaisala2}
within the field of view of TUS.
In our opinion, this is an additional argument in favor of a close relation
between slow flashes registered by TUS and lightnings.

\begin{figure}[!ht]
	\includegraphics[width=.51\textwidth]{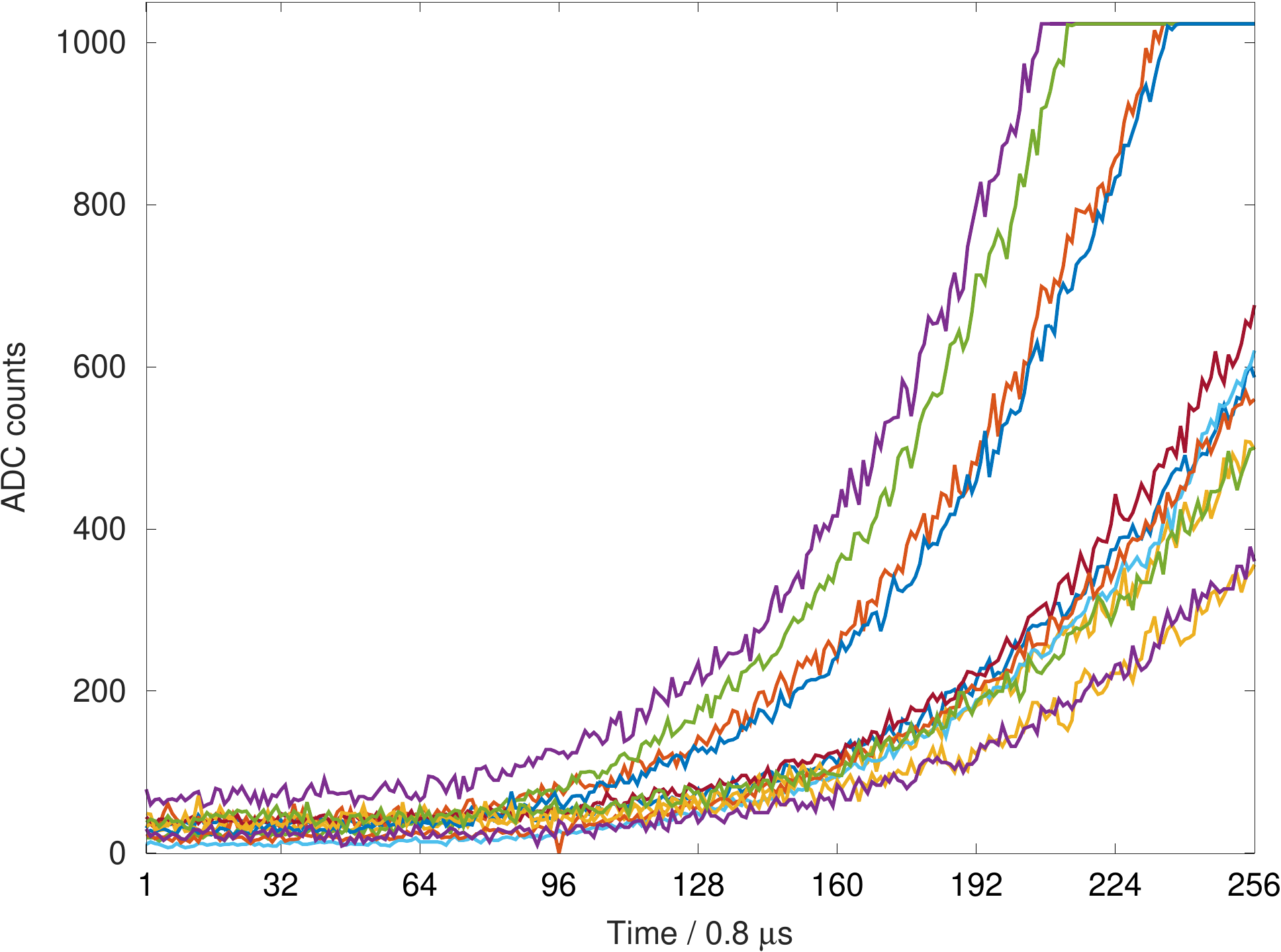}\quad
	\includegraphics[width=.47\textwidth]{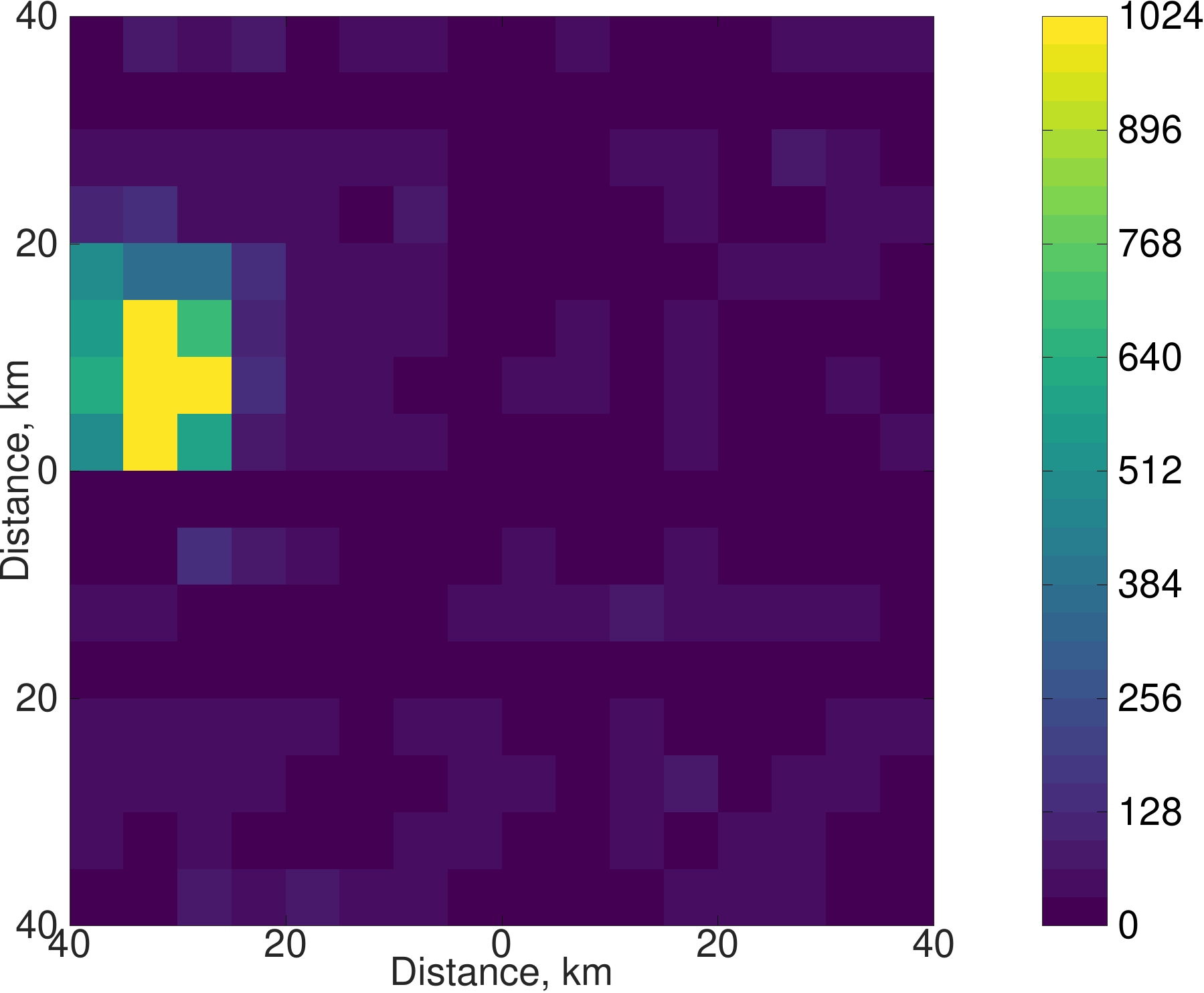}
	\caption{Extreme slow flash registered over the Black Sea on September~19, 2016.
		Left: waveforms of 12 adjacent pixels that demonstrated the
		largest growth of ADC counts.
		Right: a snapshot of the focal plane at the last frame of the
		record.
		}
	\label{fig:black_sea}
\end{figure}

There are a few events with a number of PMTs saturated before the end of
a record, which most likely represent extreme cases of slow flashes.
An example of such an event is shown in figure~\ref{fig:black_sea}.
It was registered on September~19, 2016, at 21:30:46 UTC over 
the Black Sea some 100~km to the West of Sevastopol 
($44^\circ\!\!.5$N, $32^\circ\!\!.1$E).
The event took place in conditions of high UV background
radiation. Three adjacent PMTs were saturated and another one
reached the ADC count of 1020 in about 140~$\mu$s.
The signal also increased by more than $25\sigma$ above the baseline
background level in eight neighboring pixels.  
It is known from publicly available sources%
\footnote{See, e.g., \texttt{http://en.blitzortung.org/.}} 
that a thunderstorm was taking place around the location of the TUS
event at that time, but we have failed to find information about local
lightnings registered within $\pm1$~s of the measurement.
A number of lightning strikes were collected by GLD360
and another by the WWLLN in 2--5~s after the TUS event
at distances from~20 to 50~km from its field of view. 


Another event with very similar waveforms registering as a compact spot
in the focal surface was detected at 19:54:22 UTC on September~5,
2016, near Aktobe, Kazakhstan ($50^\circ\!\!.0$N, $57^\circ\!\!.8$E),
and a similar event occurred on October~26, 2016, at 00:14:41 UTC
above the Atlantic Ocean ($31^\circ\!\!.7$N, $11^\circ\!\!.9$W).
With regard to the former, we have not managed to find information
about thunderstorms or transient luminous events nearby. 
In the latter case, the closest lightning strike was registered by 
the GLD360 approximately 2~s prior to the TUS event in $\sim400$~km
from the corresponding location.
We cannot exclude the lack of ``companions'' is due to a less
complete coverage of the two regions by lightning sensors.

There are also a few events that demonstrate compact spots in the focal
plane along with waveforms that grow monotonously, on average, but the
pixels are not saturated in any of them. 
It is likely that these represent some transitional kind of events
between the more typical slow flashes and the extreme ones, and also
relate to thunderstorms and lightnings.

Slow flashes with spot-like illumination of the focal plane
can theoretically originate from two types of rare TLEs termed
``blue jets'' and ``gigantic jets'', see, e.g.,~\cite{Pasko-jets-2008}
for a brief overview.
Both phenomena manifest themselves as optically visual
(but radiating also in UV)
jets emanating from tops of thunderclouds to the low ionosphere
but differ in their colour, typical sizes and speed of evolution.
Gigantic jets have mostly been observed above oceans, within $20^\circ$
from the equator~\cite{ISUAL-2008}.
TUS can only register the very beginning of both kinds of jets in the
EAS mode since their duration is of the order of dozens
milliseconds.\footnote{%
	TUS has registered multiple events related to thunderstorms
	in the modes with longer time samples.
	Some preliminary results can be found in~\cite{TEPA-2016}.
	}
A more thorough study is in progress.

\subsection{Events with complex spatio-temporal dynamics}

A few of the events recorded with TUS do not fit within the above three
groups but are worth discussing.

In the period of operation considered here, TUS registered three events
that are likely to be so called ``elves.''  Recall that the term
``ELVE'' stands for Emission of Light and Very Low Frequency
perturbation from an Electromagnetic Pulse (EMP)
Sources~\cite{1996GeoRL..23.2157F}.  
Elves are short-lived optical events that manifest at the lower edge of
the ionosphere (altitudes of 80--90~km) as bright rings expanding at the
speed of light up to a maximum radius of $\sim300$~km.
The life time of an elve is $\lesssim1$~ms.
They are likely caused by the atmospheric EMP ionization resulting from
cloud-to-ground lightning discharges, see,
e.g.,~\cite{1997GeoRL..24..583I}.
Elves represent the most frequent type of TLEs: according to ISUAL
(Imager of Sprites and Upper Atmospheric Lightning) global experimental
data~\cite{ISUAL-2008}, 50\% of all registered TLEs were determined to be
elves. 
The Pierre Auger collaboration performs dedicated studies of
elves~\cite{Auger-elves-ICRC2013}.

\begin{figure}[!ht]
	\includegraphics[width=0.51\textwidth]{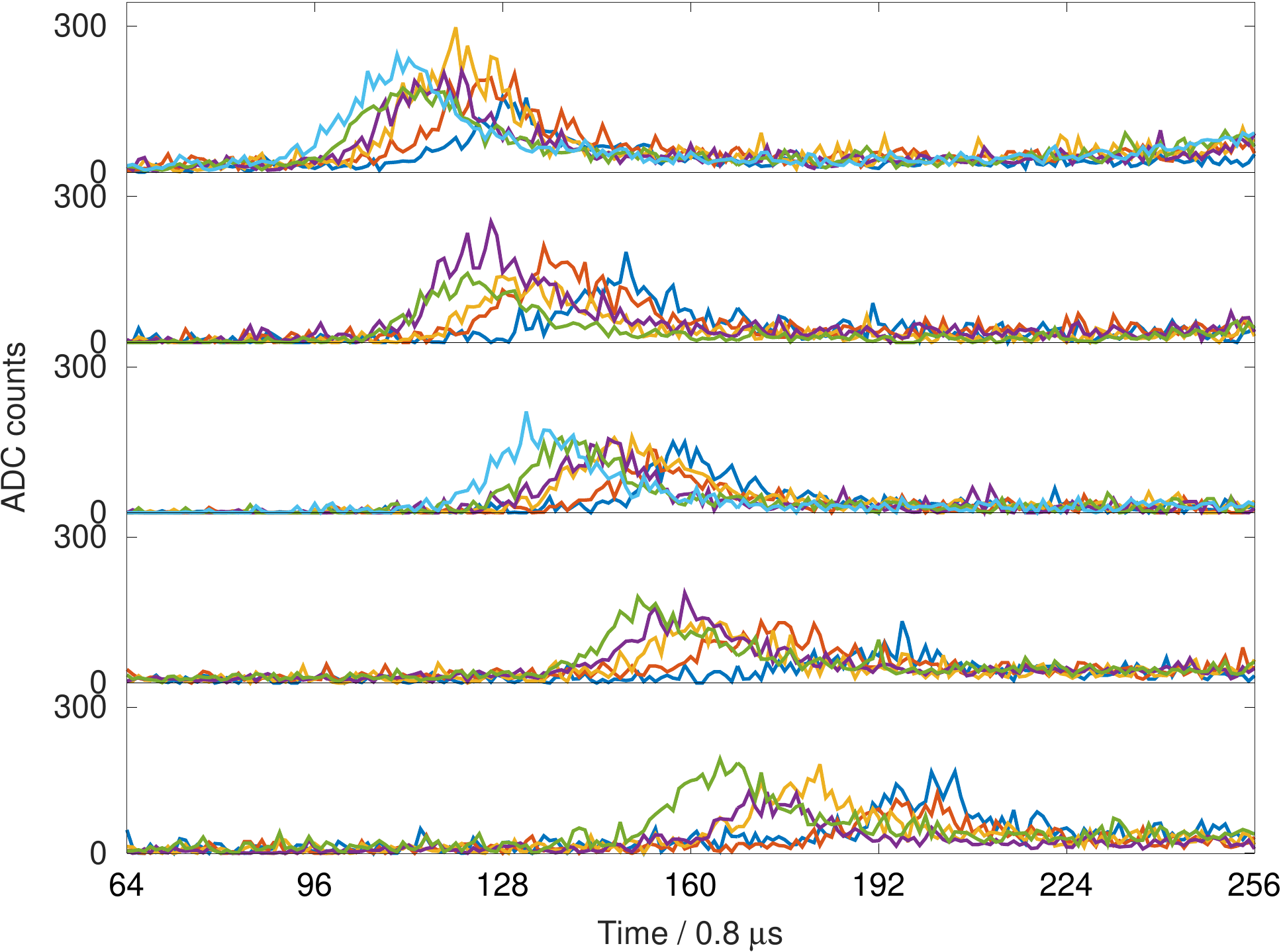}\quad
	\includegraphics[width=0.47\textwidth]{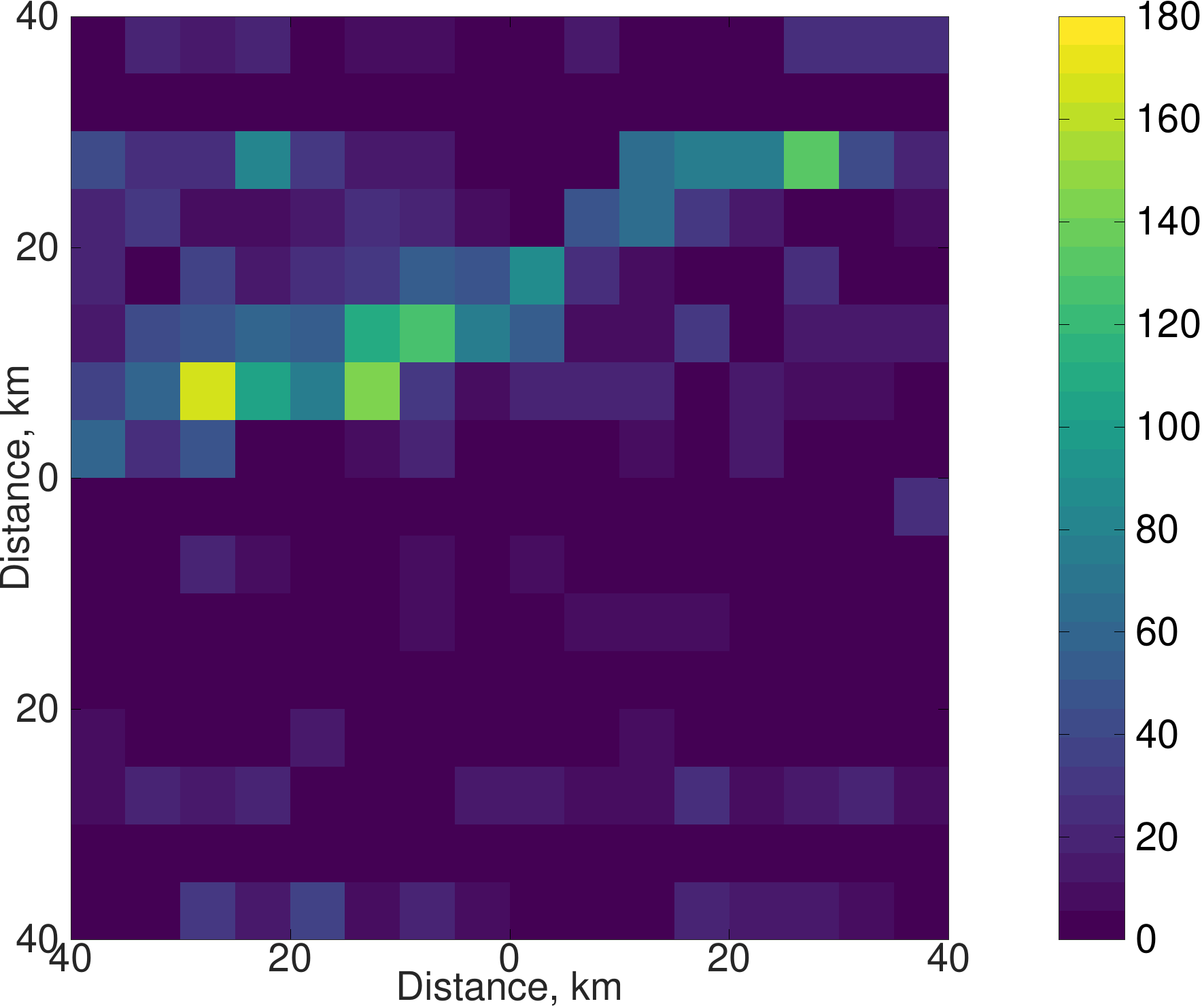}
	\caption{Event recorded over the Pacific Ocean on September~7,
		2016.
		Left: waveforms of the brightest pixels of the PMT modules shown as
		rows 3--7 from the top of the snapshot.
		Right: a snapshot of the focal plane for time frame number~172.
		The bright arc starts in the northwest corner of the
		focal plane and moves towards the southeast corner with
		decreasing brightness.
		}
	\label{fig:pacific_elve}
\end{figure}

Several waveforms and a snapshot of the focal plane of the first of the
elve candidates are shown in figure~\ref{fig:pacific_elve}.
The event was registered on September~7, 2016, at 09:51:35 UTC over the
Pacific Ocean ($11^\circ\!\!.62$S, $161^\circ\!\!.68$W) and appeared as
a bright fading arc crossing the focal plane from one corner to another
in the direction from northwest to southeast.
A simultaneous lightning (up to~1~s) was registered by the WWLLN at
a distance of about
180~km northwest from the center of the TUS field of view.
The position of the lightning, the
form of the bright arc in the focal plane, and the temporal evolution of
the waveforms strongly support the conjecture that this event was an elve.

The second event was recorded on September~18, 2016, at 22:06:48 UTC
above Chad ($9^\circ\!\!.7$N, $17^\circ\!\!.1$E).
As in the case just discussed, a bright arc crossed the focal plane from
one corner to another.
A number of lightning strikes around the location at that time were
found in GLD360, five of them occurring within $\pm1$~s of the TUS event
at distances of 100--140~km.
Their positions agree with the dynamics of the luminous arc moving
across the focal plane of TUS.
A more detailed discussion of this event can be found
in~\cite{TEPA-2016}.
The third event was recorded on October~18, 2016, at 13:20:11 UTC above
the vicinity of the Northern Mariana Islands ($15^\circ\!\!.1$N,
$149^\circ\!\!.3$E).
A lightning strike that took place in 1~s after
the TUS event at a distance of approximately 200~km
was found in the GLD360.
For both events, the UV background was similar to that of the event
shown in figure~\ref{fig:pacific_elve}.

TUS has registered several other interesting events different from
all those discussed above, but their origin is unclear and they will be
discussed elsewhere.

\section{Conclusions}

We presented preliminary results of an analysis of data obtained
by TUS, the first orbital detector of extreme energy cosmic rays,
in its main mode of operation throughout the period from August~16,
2016, to November~4, 2016.
The data have significantly improved our understanding of both the UV
background and the conditions of the space environment within which an
orbital detector of EECRs must operate,
and the diversity of events recorded in the UV range has
exceeded all expectations.

{\sloppy
Various transient UV atmospheric phenomena were studied in the earlier
MSU Tatiana~\cite{Garipov2006},
\mbox{Tatiana-2}~\cite{2010JGRA..115.0E24G} and
Vernov~\cite{2016CosRe..54..261P} satellite missions.
The instruments of these experiments included UV radiation detectors based on
the same Hamamatsu R1463 PMTs utilized in the TUS mission.
Each instrument employed a single PMT working in the ultraviolet range.
The field of view varied from $14^\circ$ to $20^\circ$ with
no spatial resolution.
Temporal resolution changed from 16~$\mu$s for the Tatiana mission
to 1~ms for the Tatiana-2.
The detectors measured the variation of the nocturnal UV glow of the
atmosphere along the satellite orbits.
The intensity of the UV glow was found to vary in a wide range
$(3$--$20)\cdot10^7$~ph~sr$^{-1}$~s$^{-1}$~cm$^{-2}$ during moonless
nights and increase up to $2\cdot10^9$~ph~sr$^{-1}$~s$^{-1}$~cm$^{-2}$
above auroral lights and during moonlit nights~\cite{2005APh....24..400G}.
Numerous transient atmospheric events were detected during the three
missions, and a global map of the nocturnal atmospheric UV glow was
composited from their data~\cite{Sadovnichy2011,2011CosRe..49..391G,Garipov2013}.

}

These studies are continuing with the TUS detector, which has a number of
crucial advantages, including its 5~km spatial resolution and the much
higher temporal resolution.  
As a result, TUS is providing new, unique information about transient
events in the atmosphere, important for the development of much more
advanced instruments, such as KLYPVE (\mbox{K-EUSO}) and JEM-EUSO.
TUS continues its orbital mission, and a detailed analysis of its data
is in progress, including a dedicated search for EASs generated by
cosmic rays of the highest energies.
No events of this kind have been unequivocally identified in the data yet,
but a preliminary list of possible candidates is under detailed
analysis~\cite{TUS-candidates-arxiv}.
We believe the experiment marks an important
step in the development of the technique of detecting extreme energy
cosmic rays from space.

\acknowledgments

The authors wish to thank Robert~Holzworth, the head of
the World Wide Lightning Location Network,
and Vaisala Inc.\ company for providing the data on lightning strikes 
employed in the present study.
We also thank an anonymous referee for multiple useful comments on the
manuscript.
The work was done with partial financial support from the Russian
Foundation for Basic Research grants No.\ 15-02-05498-a and No.\ 16-29-13065.
The Korean work is supported by the National Research Foundation grants
(No.\ 2015R1A2A1A01006870 and No.\ 2015R1A2A1A15055344).

\bibliographystyle{JHEP}
\bibliography{tus_first_v2}

\providecommand{\href}[2]{#2}\begingroup\raggedright\begin{thebibliography}{10}

\bibitem{1961PhRvL...6..485L}
J.~{Linsley}, L.~{Scarsi} and B.~{Rossi}, \emph{{Extremely Energetic Cosmic-Ray
  Event}}, \href{http://dx.doi.org/10.1103/PhysRevLett.6.485}{\emph{Physical
  Review Letters} {\bfseries 6} (May, 1961) 485--487}.

\bibitem{Kampert-Tinyakov-2014}
K.-H. {Kampert} and P.~{Tinyakov}, \emph{Cosmic rays from the ankle to the
  cutoff}, \href{http://dx.doi.org/10.1016/j.crhy.2014.04.006}{\emph{Comptes
  Rendus Physique} {\bfseries 15} (2014) 318--328},
  [\href{https://arxiv.org/abs/1405.0575}{{\ttfamily 1405.0575}}].

\bibitem{2015ApJ...804...15A}
A.~{Aab}, P.~{Abreu}, M.~{Aglietta} et~al., \emph{{Searches for anisotropies in
  the arrival directions of the highest energy cosmic rays detected by the
  Pierre Auger Observatory}},
  \href{http://dx.doi.org/10.1088/0004-637X/804/1/15}{\emph{Astrophys. J.}
  {\bfseries 804} (2015) 15},
  [\href{https://arxiv.org/abs/1411.6111}{{\ttfamily 1411.6111}}].

\bibitem{Auger-TA-2015}
{\scshape Pierre Auger, Telescope Array} collaboration, R.~U. {Abbasi},
  M.~{Abe}, T.~{Abu-Zayyad} et~al., \emph{{Pierre Auger Observatory and
  Telescope Array: Joint Contributions to the 34th International Cosmic Ray
  Conference (ICRC 2015)}},  \href{https://arxiv.org/abs/1511.02103}{{\ttfamily
  1511.02103}}.

\bibitem{1980BAAS...12Q.818B}
R.~{Benson} and J.~{Linsley}, \emph{{Satellite observation of cosmic-ray air
  showers}},  in \emph{Bulletin of the American Astronomical Society}, vol.~12,
  p.~818, 1980.

\bibitem{Benson-Linsley-1981}
R.~{Benson} and J.~{Linsley}, \emph{Satellite observation of cosmic ray air
  showers},  in \emph{17th International Cosmic Ray Conference, Paris, France},
  vol.~8, pp.~145--148, 1981.

\bibitem{ECRS}
M.~I. {Panasyuk}, M.~{Casolino}, G.~K. {Garipov} et~al., \emph{{The current
  status of orbital experiments for UHECR studies}},
  \href{http://dx.doi.org/10.1088/1742-6596/632/1/012097}{\emph{Journal of
  Physics Conference Series} {\bfseries 632} (2015) 012097},
  [\href{https://arxiv.org/abs/1501.06368}{{\ttfamily 1501.06368}}].

\bibitem{2001ICRC....2..831A}
V.~V. {Alexandrov}, D.~I. {Bugrov}, G.~K. {Garipov} et~al., \emph{{Space
  experiment ``TUS'' for study of ultra high energy cosmic rays}},
  {\emph{International Cosmic Ray Conference} {\bfseries 2} (2001) 831}.

\bibitem{2001AIPC..566...57K}
B.~A. {Khrenov}, M.~I. {Panasyuk}, V.~V. {Alexandrov} et~al., \emph{{Space
  Program KOSMOTEPETL (project KLYPVE and TUS) for the study of extremely high
  energy cosmic rays}},  in \emph{Observing Ultrahigh Energy Cosmic Rays from
  Space and Earth} (H.~{Salazar}, L.~{Villasenor} and A.~{Zepeda}, eds.),
  vol.~566 of \emph{American Institute of Physics Conference Series},
  pp.~57--75, 2001.
\newblock \href{http://dx.doi.org/10.1063/1.1378622}{DOI}.

\bibitem{2004PAN....67.2058K}
B.~A. {Khrenov}, V.~V. {Alexandrov}, D.~I. {Bugrov} et~al., \emph{{KLYPVE/TUS
  space experiments for study of ultrahigh-energy cosmic rays}},
  \href{http://dx.doi.org/10.1134/1.1825529}{\emph{Physics of Atomic Nuclei}
  {\bfseries 67} (2004) 2058--2061}.

\bibitem{2008AdSpR..41.2079A}
V.~{Abrashkin}, V.~{Alexandrov}, Y.~{Arakcheev} et~al., \emph{{Updated TUS
  space fluorescence detector for study of UHECR}},
  \href{http://dx.doi.org/10.1016/j.asr.2007.03.036}{\emph{Advances in Space
  Research} {\bfseries 41} (2008) 2079--2088}.

\bibitem{TUS-ecrs2012}
B.~A. {Khrenov}, M.~I. {Panasyuk}, G.~K. {Garipov} et~al., \emph{{Pioneering
  space based detector for study of cosmic rays beyond GZK limit}},  in
  \emph{European Physical Journal Web of Conferences}, vol.~53, p.~09006, 2013.
\newblock \href{http://dx.doi.org/10.1051/epjconf/20135309006}{DOI}.

\bibitem{Lomonosov2013}
V.~A. Sadovnichiy, A.~M. Amelyushkin, V.~Angelopoulos et~al., \emph{{Space
  experiments aboard the Lomonosov MSU satellite}},
  \href{http://dx.doi.org/10.1134/S0010952513060063}{\emph{Cosmic Research}
  {\bfseries 51} (2013) 427--433}.

\bibitem{klypve-2015}
G.~K. {Garipov}, M.~Y. {Zotov}, P.~A. {Klimov} et~al., \emph{The {KLYPVE} ultra
  high energy cosmic ray detector on board the {ISS}},
  \href{http://dx.doi.org/10.3103/S1062873815030193}{\emph{Bull. Rus. Acad.
  Sci. Physics} {\bfseries 79} (2015) 326--328}.

\bibitem{JEM-EUSO-intro}
J.~{Adams Jr.}, S.~Ahmad, J.-N. Albert et~al., \emph{{The JEM-EUSO mission: An
  introduction}},
  \href{http://dx.doi.org/10.1007/s10686-015-9482-x}{\emph{Experimental
  Astronomy} {\bfseries 40} (2015) 3--17}.

\bibitem{JEM-EUSO-tool}
J.~{Adams Jr.}, S.~Ahmad, J.-N. Albert et~al., \emph{{The JEM-EUSO
  instrument}},
  \href{http://dx.doi.org/10.1007/s10686-014-9418-x}{\emph{Experimental
  Astronomy} {\bfseries 40} (2015) 19--44}.

\bibitem{Olinto-ICRC-2015}
A.~{Olinto}, E.~{Parizot}, M.~{Bertaina} and G.~{Medina-Tanco}, \emph{{JEM-EUSO
  Science}}, {\emph{Proceedings of Science (ICRC2015)} (2015) 623}.

\bibitem{Semikoz:2016mfg}
D.~Semikoz, P.~Tinyakov and M.~Zotov, \emph{{Detection prospects of the
  Telescope Array hotspot by space observatories}},
  \href{http://dx.doi.org/10.1103/PhysRevD.93.103005}{\emph{Phys. Rev.}
  {\bfseries D93} (2016) 103005},
  [\href{https://arxiv.org/abs/1601.06363}{{\ttfamily 1601.06363}}].

\bibitem{2011ICRC....3..124T}
L.~{Tkachev}, A.~{Grinyuk}, V.~{Grebenyuk} et~al., \emph{{The TUS Fresnel
  mirror production and optical parameters measurement}},
  \href{http://dx.doi.org/10.7529/ICRC2011/V03/0659}{\emph{International Cosmic
  Ray Conference} {\bfseries 3} (2011) 124}.

\bibitem{2014NIMPA.763..604G}
A.~{Grinyuk}, M.~{Slunecka}, A.~{Tkachenko} et~al., \emph{{The method and
  results of measurement of the optical parameters of the UHECR detector for
  the TUS space experiment}},
  \href{http://dx.doi.org/10.1016/j.nima.2014.06.019}{\emph{Nuclear Instruments
  and Methods in Physics Research A} {\bfseries 763} (2014) 604--609}.

\bibitem{SSR}
P.~A. {Klimov}, M.~I. {Panasyuk}, B.~A. {Krenov} and {Lomonosov--UHECR/TLE
  Collaboration}, \emph{The {TUS} detector of extreme energy cosmic rays on
  board the {Lomonosov} satellite}, {\emph{ArXiv e-prints} (2017) },
  [\href{https://arxiv.org/abs/1706.04976}{{\ttfamily 1706.04976}}].

\bibitem{2011ICRC...11..272T}
A.~{Tkachenko}, V.~{Boreiko}, G.~{Garipov} et~al., \emph{{Photo receiver of the
  orbital ultra high energy cosmic rays detector TUS at Earth and in planetary
  environments weather}},
  \href{http://dx.doi.org/10.7529/ICRC2011/V11/1231}{\emph{International Cosmic
  Ray Conference} {\bfseries 11} (2011) 272}.

\bibitem{ESAF2010}
C.~{Berat}, S.~{Bottai}, D.~{De Marco} et~al., \emph{Full simulation of
  space-based extensive air showers detectors with {ESAF}},
  \href{http://dx.doi.org/10.1016/j.astropartphys.2010.02.005}{\emph{Astroparticle
  Physics} {\bfseries 33} (2010) 221--247},
  [\href{https://arxiv.org/abs/0907.5275}{{\ttfamily 0907.5275}}].

\bibitem{2013JPhCS.409a2105G}
A.~A. {Grinyuk}, A.~V. {Tkachenko} and L.~G. {Tkachev}, \emph{{The TUS orbital
  detector optical system and trigger simulation}},
  \href{http://dx.doi.org/10.1088/1742-6596/409/1/012105}{\emph{Journal of
  Physics Conference Series} {\bfseries 409} (2013) 012105}.

\bibitem{sim-icrc2015}
L.~{Tkachev}, A.~{Grinyuk}, M.~{Lavrova} and A.~{Tkachenko}, \emph{The {TUS}
  orbital detector simulation}, {\emph{Proceedings of Science} (2015) 610}.

\bibitem{TUS-sim-ApP}
A.~{Grinyuk}, V.~{Grebenyuk}, B.~{Khrenov} et~al., \emph{The orbital {TUS}
  detector simulation},
  \href{http://dx.doi.org/10.1016/j.astropartphys.2016.09.003}{\emph{Astroparticle
  Physics} {\bfseries 90} (2017) 93--97}.

\bibitem{TUS-expastron}
J.~{Adams Jr.}, S.~Ahmad, J.-N. Albert et~al., \emph{{Space experiment TUS on
  board the Lomonosov satellite as pathfinder of JEM-EUSO}},
  \href{http://dx.doi.org/10.1007/s10686-015-9465-y}{\emph{Experimental
  Astronomy} {\bfseries 40} (2015) 315--326}.

\bibitem{Auger-spectrum-2015}
{The Pierre Auger Collaboration}, \emph{Measurement of the cosmic ray spectrum
  above $4\times10^{18}$ ev using inclined events detected with the {P}ierre
  {A}uger {O}bservatory},
  \href{http://dx.doi.org/10.1088/1475-7516/2015/08/049}{\emph{Journal of
  Cosmology and Astroparticle Physics} {\bfseries 8} (2015) 049}.

\bibitem{uhecr2016}
{\scshape Lomonosov-UHECR/TLE} collaboration, M.~{Zotov}, \emph{Early results
  from {TUS}, the first orbital detector of extreme energy cosmic rays},
  {\emph{ArXiv e-prints} (Mar., 2017) },
  [\href{https://arxiv.org/abs/1703.09484}{{\ttfamily 1703.09484}}].

\bibitem{2016SciA....2E0377F}
F.~{Falchi}, P.~{Cinzano}, D.~{Duriscoe} et~al., \emph{{The new world atlas of
  artificial night sky brightness}},
  \href{http://dx.doi.org/10.1126/sciadv.1600377}{\emph{Science Advances}
  {\bfseries 2} (2016) e1600377--e1600377},
  [\href{https://arxiv.org/abs/1609.01041}{{\ttfamily 1609.01041}}].

\bibitem{tracks-izvran}
P.~A. {Klimov}, M.~Y. {Zotov}, N.~P. {Chirskaya} et~al., \emph{Preliminary
  results of the {TUS} orbital telescope of ultra-high energy cosmic rays:
  registration of low-energy particles passing through the photodetector},
  \href{http://dx.doi.org/10.3103/S1062873817040256}{\emph{Bull. Rus. Acad.
  Sci. Physics} {\bfseries 81} (2017) 407--409}.

\bibitem{ecrs2016}
S.~V. {Biktemerova}, A.~V. {Bogomolov}, V.~V. {Bogomolov} et~al., \emph{First
  results of the {L}omonosov {TUS} and {GRB} experiments},  in \emph{{25th
  European Cosmic Ray Symposium (ECRS 2016) Turin, Italy, September 04-09,
  2016}}, 2017.
\newblock \href{https://arxiv.org/abs/1703.03738}{{\ttfamily 1703.03738}}.

\bibitem{Geant4}
S.~{Agostinelli}, J.~{Allison}, K.~{Amako} et~al., \emph{{GEANT4---a simulation
  toolkit}},
  \href{http://dx.doi.org/10.1016/S0168-9002(03)01368-8}{\emph{Nuclear
  Instruments and Methods in Physics Research A} {\bfseries 506} (2003)
  250--303}.

\bibitem{2011BRASP..75..381S}
O.~P. {Shustova}, N.~N. {Kalmykov} and B.~A. {Khrenov}, \emph{{Possibility of
  using satellite-based detector for recording the Cherenkov light from
  ultrahigh-energy extensive atmospheric shower penetrating into ocean water}},
  \href{http://dx.doi.org/10.3103/S1062873811030385}{\emph{Bull. Russ. Acad.
  Sci., Physics} {\bfseries 75} (2011) 381--384},
  [\href{https://arxiv.org/abs/1110.2974}{{\ttfamily 1110.2974}}].

\bibitem{2014AdSpR..53.1515B}
M.~{Bertaina}, S.~{Biktemerova}, K.~{Bittermann} et~al., \emph{{Performance and
  air-shower reconstruction techniques for the JEM-EUSO mission}},
  \href{http://dx.doi.org/10.1016/j.asr.2014.02.018}{\emph{Advances in Space
  Research} {\bfseries 53} (2014) 1515--1535}.

\bibitem{vaisala1}
R.~K. {Said}, U.~S. {Inan} and K.~L. {Cummins}, \emph{{Long-range lightning
  geolocation using a VLF radio atmospheric waveform bank}},
  \href{http://dx.doi.org/10.1029/2010JD013863}{\emph{Journal of Geophysical
  Research (Atmospheres)} {\bfseries 115} (2010) D23108}.

\bibitem{vaisala2}
R.~K. {Said}, M.~B. {Cohen} and U.~S. {Inan}, \emph{{Highly intense lightning
  over the oceans: Estimated peak currents from global GLD360 observations}},
  \href{http://dx.doi.org/10.1002/jgrd.50508}{\emph{Journal of Geophysical
  Research (Atmospheres)} {\bfseries 118} (2013) 6905--6915}.

\bibitem{Pasko-jets-2008}
V.~P. {Pasko}, \emph{Blue jets and gigantic jets: transient luminous events
  between thunderstorm tops and the lower ionosphere}, {\emph{Plasma Physics
  and Controlled Fusion} {\bfseries 50} (2008) 124050}.

\bibitem{ISUAL-2008}
A.~B. {Chen}, C.-L. {Kuo}, Y.-J. {Lee} et~al., \emph{Global distributions and
  occurrence rates of transient luminous events},
  \href{http://dx.doi.org/10.1029/2008JA013101}{\emph{Journal of Geophysical
  Research: Space Physics} {\bfseries 113} (2008) n/a--n/a}.

\bibitem{TEPA-2016}
P.~{Klimov}, B.~{Khrenov}, S.~{Sharakin} et~al., \emph{First results on
  transient atmospheric events from {T}racking {U}ltraviolet {S}etup ({TUS}) on
  board the {L}omonosov satellite},  in \emph{Proc. of {I}nt. {S}ymposium
  {TEPA}-2016} (A.~{Chilingarian}, ed.), pp.~122--127, 2017.

\bibitem{1996GeoRL..23.2157F}
H.~{Fukunishi}, Y.~{Takahashi}, M.~{Kubota} et~al., \emph{{Elves:
  Lightning-induced transient luminous events in the lower ionosphere}},
  \href{http://dx.doi.org/10.1029/96GL01979}{\emph{Geophysical Research
  Letters} {\bfseries 23} (1996) 2157--2160}.

\bibitem{1997GeoRL..24..583I}
U.~S. {Inan}, C.~{Barrington-Leigh}, S.~{Hansen} et~al., \emph{{Rapid lateral
  expansion of optical luminosity in lightning-induced ionospheric flashes
  referred to as ``elves''}},
  \href{http://dx.doi.org/10.1029/97GL00404}{\emph{Geophysical Research
  Letters} {\bfseries 24} (1997) 583--586}.

\bibitem{Auger-elves-ICRC2013}
{\scshape Pierre Auger} collaboration, A.~{Tonachini}, \emph{Observation of
  elves at the {P}ierre {A}uger {O}bservatory},  in \emph{{Proceedings, 33rd
  International Cosmic Ray Conference (ICRC2013): Rio de Janeiro, Brazil, July
  2-9, 2013}}, 2013.
\newblock \href{https://arxiv.org/abs/1307.5059}{{\ttfamily 1307.5059}}.

\bibitem{Garipov2006}
G.~K. Garipov, M.~I. Panasyuk, I.~A. Rubinshtein et~al., \emph{Ultraviolet
  radiation detector of the {MSU} research educational microsatellite
  {U}niversitetskii-{T}at'yana},
  \href{http://dx.doi.org/10.1134/S0020441206010180}{\emph{Instruments and
  Experimental Techniques} {\bfseries 49} (2006) 126--131}.

\bibitem{2010JGRA..115.0E24G}
G.~K. {Garipov}, B.~A. {Khrenov}, P.~A. {Klimov} et~al., \emph{{Program of
  transient UV event research at Tatiana-2 satellite}},
  \href{http://dx.doi.org/10.1029/2009JA014765}{\emph{Journal of Geophysical
  Research (Space Physics)} {\bfseries 115} (2010) A00E24}.

\bibitem{2016CosRe..54..261P}
M.~I. {Panasyuk}, S.~I. {Svertilov}, V.~V. {Bogomolov} et~al.,
  \emph{{Experiment on the Vernov satellite: Transient energetic processes in
  the Earth's atmosphere and magnetosphere. Part I: Description of the
  experiment}}, \href{http://dx.doi.org/10.1134/S0010952516040043}{\emph{Cosmic
  Research} {\bfseries 54} (2016) 261--269}.

\bibitem{2005APh....24..400G}
G.~K. {Garipov}, B.~A. {Khrenov}, M.~I. {Panasyuk} et~al., \emph{{UV radiation
  from the atmosphere: Results of the MSU ``Tatiana'' satellite measurements}},
  \href{http://dx.doi.org/10.1016/j.astropartphys.2005.09.001}{\emph{Astroparticle
  Physics} {\bfseries 24} (2005) 400--408}.

\bibitem{Sadovnichy2011}
V.~A. Sadovnichy, M.~I. Panasyuk, I.~V. Yashin et~al., \emph{{Investigations of
  the space environment aboard the Universitetsky-Tat'yana and
  Universitetsky-Tat'yana-2 microsatellites}},
  \href{http://dx.doi.org/10.1134/S0038094611010096}{\emph{Solar System
  Research} {\bfseries 45} (2011) 3--29}.

\bibitem{2011CosRe..49..391G}
G.~K. {Garipov}, P.~A. {Klimov}, V.~S. {Morozenko} et~al., \emph{{Time and
  energy characteristics of UV flashes in the atmosphere: Data of the
  Universitetsky-Tatiana satellite}},
  \href{http://dx.doi.org/10.1134/S0010952511050066}{\emph{Cosmic Research}
  {\bfseries 49} (2011) 391--398}.

\bibitem{Garipov2013}
G.~K. Garipov, B.~A. Khrenov, P.~A. Klimov et~al., \emph{{Global transients in
  ultraviolet and red-infrared ranges from data of Universitetsky-Tatiana-2
  satellite}}, \href{http://dx.doi.org/10.1029/2012JD017501}{\emph{Journal of
  Geophysical Research: Atmospheres} {\bfseries 118} (2013) 370--379}.

\bibitem{TUS-candidates-arxiv}
S.~V. {Biktemerova}, A.~A. {Botvinko}, N.~P. {Chirskaya} et~al., \emph{Search
  for extreme energy cosmic ray candidates in the {TUS} orbital experiment
  data}, {\emph{ArXiv e-prints} (June, 2017) },
  [\href{https://arxiv.org/abs/1706.05369}{{\ttfamily 1706.05369}}].

\end{thebibliography}\endgroup

\end{document}